\pdfoutput=1
\documentclass[aps,prx,reprint,superscriptaddress,amsmath,amssymb,longbibliography,floatfix]{revtex4-2}
\usepackage[T1]{fontenc}
\usepackage{amsmath,amssymb,amsthm,bm,mathtools}
\usepackage{graphicx,xcolor,booktabs,array}
\usepackage{tikz}
\usetikzlibrary{arrows.meta,positioning}
\tikzset{sd/.style={font=\scriptsize,>=Latex},
 tb/.style={draw,minimum size=4.5mm,inner sep=0pt},
 op/.style={tb,fill=orange!30,draw=orange!80!black}}
\usepackage{hyperref}
\hypersetup{colorlinks=true,linkcolor=blue!45!black,citecolor=blue!45!black,
 urlcolor=blue!45!black,pdftitle={Transfer-Matrix Calculus for Infinite Matrix Product States},
 pdfauthor={Qi Yang and Philippe Corboz}}
\graphicspath{{figures/}}
\newcommand{\id}{I}
\newcommand{\ii}{\mathrm{i}}
\newcommand{\dd}{\mathrm{d}}
\newcommand{\e}{\mathrm{e}}
\newcommand{\ket}[1]{\lvert #1\rangle}
\newcommand{\bra}[1]{\langle #1\rvert}
\newcommand{\braket}[2]{\langle #1\vert #2\rangle}
\newcommand{\Li}{\operatorname{Li}}

\newcommand{\ddf}[1]{F_J^{[#1]}}
\newcommand{\Qproj}{P}%
\newcommand{\Tconn}{\widetilde T}
\newcommand{\slot}{\mathord{\vcenter{\hbox{\scalebox{1.6}{$\cdot$}}}}}
\newcommand{\gapmax}{q_{\max}}
\newcommand{\Uone}{\mathrm{U}(1)}

\begin{document}
\title{Transfer-Matrix Calculus for Infinite Matrix Product States}
\author{Qi Yang}
\email{qiyang@mail.ustc.edu.cn}
\affiliation{Institute for Theoretical Physics, University of Amsterdam,
1098 XH Amsterdam, The Netherlands}
\author{Philippe Corboz}
\affiliation{Institute for Theoretical Physics, University of Amsterdam,
1098 XH Amsterdam, The Netherlands}
\date{\today}

\begin{abstract}
Long-range interactions, from the Coulomb repulsion between electrons to the power-law interactions of trapped ions and Rydberg atoms, are widespread in quantum many-body physics. For infinite systems, however, matrix product state (MPS) algorithms, the standard classical reference in one dimension, conventionally rely on a computational compromise: replacing these interactions by fitted sums of exponentials.
We develop transfer-matrix calculus (TMC), a unified framework for deriving infinite-MPS algorithms. In TMC, an interaction is written as a function of the transfer matrix, and one set of contraction rules then yields the effective Hamiltonians of the density-matrix renormalization group method and the variational uniform matrix product state method, the tangent-space excitation action, and the energy
variance, without a matrix product operator (MPO). Long-range interactions thus enter with their prescribed distance dependence, without a fitted sum of exponentials.
For the half-filled lattice Coulomb chain with a $1/d$ interaction of strength $V$, theory predicts charge-density-wave (CDW) order for all $V>0$, but at weak coupling this order is too weak for earlier numerics to resolve. Using TMC, we obtain infinite-MPS ground states at $V=2$ directly in the thermodynamic limit, with an energy variance per site of $2\times10^{-10}$. Our data, extrapolated to zero energy variance, indicate a small nonzero CDW amplitude, $m_0=0.002213(14)$, consistent with this prediction. The complete $V=2$ optimization workflow across the full series of bond dimensions takes less than a day on a single AMD MI250X GPU die.
For an exactly solvable long-range hopping model, TMC reproduces the nonanalytic $|k|$ cusp of the one-particle dispersion that any fixed finite sum of decaying exponentials rounds off.
We also use TMC to optimize first-quantized MPS directly in the thermodynamic limit. For half-filled free fermions, our entanglement entropy data show scaling consistent with central charge $c=1$ over the studied bond dimensions, as for second-quantized MPS.
\end{abstract}
\maketitle

\section{Introduction}
\label{sec:introduction}
Long-range interactions are ubiquitous in nature and arise in both field theory and quantum simulators. In field theory, a sufficiently slowly decaying power-law tail can change the universality class relative to the short-range case, an effect studied in work ranging from the early renormalization-group analyses of long-range ferromagnets~\cite{Fisher1972,Sak1973} to recent studies of long-range quantum matter~\cite{Defenu2023,Schuckert2020}. Quantum simulators based on trapped ions~\cite{Richerme2014,Jurcevic2014}, Rydberg arrays~\cite{Bernien2017,Saffman2010}, polar molecules~\cite{Yan2013}, and magnetic atoms realize couplings that decay as a power of the distance~\cite{Defenu2023}; in trapped ions the power-law form is approximate and its exponent tunable~\cite{Monroe2021}. Benchmarking these simulators requires comparing their measurements with classical reference results for the long-range models they implement~\cite{Hauke2012,Daley2022,Joshi2022}. As accessible system sizes increase, this need motivates calculations of thermodynamic-limit ground states and excitation spectra, together with diagnostics of state convergence.

The lattice chain with repulsive $1/d$ Coulomb interactions and a neutralizing background provides a concrete test of such reference calculations.
Theory predicts that the ground state of the half-filled lattice Coulomb chain develops charge order at every nonzero repulsive coupling. Using bosonization, Schulz showed that the $1/d$ interaction produces quasi-Wigner-crystal correlations in a one-dimensional electron gas in the continuum~\cite{Schulz1993}. In the half-filled lattice chain, commensurability should pin a charge-density wave (CDW), although with an extremely small amplitude at weak coupling~\cite{Capponi2000,Valenzuela2003}.
At weak coupling, numerics have not detected this order. Capponi \emph{et al.}\ agreed with this field-theory prediction but could not resolve the order numerically: in rings of up to 28 sites they found a quasimetallic regime~\cite{Capponi2000}. The more recent phase diagram of Ref.~\cite{Li2019} places weak coupling on the metallic side.
Resolving such weak order requires high resolution and controlled variational extrapolation with the exact interaction.

Matrix product states (MPS) provide a variational framework for studying one-dimensional quantum many-body systems. They underpin the density-matrix renormalization group (DMRG)~\cite{White1992,Fannes1992,Ostlund1995,Schollwock2011,Cirac2021}, and uniform infinite MPS (iMPS) represent translation-invariant states directly in the thermodynamic limit~\cite{Vidal2007,McCulloch2008}. Variational and tangent-space formulations give access to ground states, excitations, and dynamics~\cite{Haegeman2011TDVP,Haegeman2013Tangent,Haegeman2016,Zauner2018,Vanderstraeten2019}. In an iMPS, the transfer matrix plays a central role: its fixed points are the left and right environments, and its subleading modes govern the spatial decay of correlations~\cite{Fannes1992,McCulloch2008,Zauner2015,HaegemanVerstraete2017}. However, standard variational optimization is not usually formulated directly in terms of the iMPS transfer matrix.

Infinite DMRG (iDMRG) and the variational uniform MPS (VUMPS) method optimize iMPS ground states using effective Hamiltonians~\cite{McCulloch2008,Zauner2018}. For nearest-neighbor and other finite-range interactions, or finite sums of decaying exponentials, the Hamiltonian has an exact matrix product operator (MPO) representation of finite bond dimension~\cite{Crosswhite2008,Pirvu2010,Schollwock2011}. MPO contractions supply the environments and effective Hamiltonians for these algorithms. A decaying power-law interaction on an infinite chain has no such exact representation, so the MPO approach typically uses a fitted sum of exponentials (SOE)~\cite{Crosswhite2008,Pirvu2010,Frowis2010,BeylkinMonzon2010}. The number of exponentials and the fitting range must be converged alongside the MPS bond dimension. A fixed decaying SOE replaces the algebraic tail by an exponential one, adding Hamiltonian error to the variational error in ground-state, excitation, and variance calculations.

In this work we develop the \emph{transfer-matrix calculus} (TMC) to compute iMPS ground states, excitation spectra, and energy variances for algebraically decaying interactions without an SOE fit. The interaction
enters as a function of the transfer matrix. Our earlier work used this representation to evaluate the energy and its gradient~\cite{Yang2026}. Here a common set of contraction rules yields the effective Hamiltonians required by iDMRG and VUMPS, the tangent-space excitation action, and the full energy variance. For the energy, we show that these rules reproduce the MPO contraction identically whenever an exact MPO representation of the Hamiltonian exists, so the standard MPO approach is a special case.

TMC builds on two existing ideas: differentiable tensor networks~\cite{Liao2019} and generating functions for sums of diagrams~\cite{Tu2021,Tu2024}. In differentiable tensor-network methods, the derivative of a contraction with respect to one tensor is obtained by removing that tensor and leaving its indices open. The action of the effective Hamiltonian in a variational algorithm is such a derivative of the energy, and excitation actions are obtained in the same way~\cite{Ponsioen2022}. In the generating-function approach, one function collects a sum over diagrams, and its derivatives collect tensor insertions. Automatic differentiation obtains such derivatives by differentiating every step of a computation. TMC instead first sums the interaction over all distances into a single function of the transfer matrix, namely the generating function of the interaction, and then differentiates this function as a whole. 

We demonstrate the accuracy and versatility of TMC through three complementary applications. First, we apply TMC to the ground state of the half-filled Coulomb chain with the exact $1/d$ interaction. Extrapolating to zero energy variance, we obtain a high-precision estimate of the CDW amplitude $0.09923211059(3)$ at interaction strength $V=4$. At $V=2$, where earlier numerics placed the chain on the metallic side~\cite{Li2019}, our data indicate a small nonzero CDW amplitude of $0.002213(14)$, consistent with the theoretical prediction.

We next apply TMC to excitations. For an exactly solvable model with inverse-square hopping, we reproduce the nonanalytic $|k|$ cusp of the one-particle dispersion to machine precision over six decades of momentum. In contrast, approximating the hopping by a finite sum of exponentials yields a dispersion that is analytic at $k=0$, rounding off the cusp, even when the fit is highly accurate over a large but finite range of distances.

Finally, we extend TMC to first-quantized MPS in the thermodynamic limit. Following the construction of Li and Waintal~\cite{LiWaintal2026}, each tensor is associated with a particle, and the physical indices encode the separations between neighboring particles. Unlike in second-quantized MPS, the number of transfer matrices separating two particles does not encode their spatial distance. Nevertheless, we show that the interaction can still be expressed as a function of the first-quantized iMPS transfer matrix, enabling variational optimization without an SOE fit. For the half-filled Coulomb chain, the scaling of the entanglement entropy at $V=0$ is consistent with central charge $c=1$, and at nonzero coupling first- and second-quantized states fall on one curve of CDW error against energy error.

This paper is organized as follows. In Sec.~\ref{sec:calculus} we develop the TMC framework and algorithms. Sections~\ref{sec:coulomb}, \ref{sec:excitations}, and \ref{sec:firstq} apply TMC to the ground state of the Coulomb chain, to excitations, and to first-quantized iMPS, respectively. Section~\ref{sec:discussion} summarizes our main findings and outlines potential applications and extensions of TMC. %
\section{Transfer-matrix calculus}
\label{sec:calculus}
\label{sec:effective}

In this section, we introduce TMC for Hamiltonians with pairwise interactions. In the MPS energy evaluation, each interaction term connects two sites by a string of transfer matrices and is weighted by the interaction strength at their separation. Summing over the interaction distance makes the energy a function of the transfer matrix. The gradient, the effective Hamiltonians of iDMRG and VUMPS, and the excitation action leave one or two tensors in these strings open and sum over their positions. Section~\ref{sec:open-sums} collects the results on derivatives of matrix functions that these sums require. Sections~\ref{sec:energy-gradient}, \ref{sec:effective-operators}, and~\ref{sec:variance-method} apply them to the energy and gradient, the effective Hamiltonians, and the energy variance; the excitation action follows in Sec.~\ref{sec:excitations}. 

\subsection{Derivatives of matrix functions}
\label{sec:open-sums}

In this subsection, we collect the results on derivatives of matrix functions that the rest of the paper uses. These results are standard in the mathematical literature~\cite{Mathias1996,Higham2008}. We consider a matrix function $f(T)=\sum_{n\geq0}c_nT^n$, defined by a polynomial or power series that converges on the spectrum of $T$. We first derive the first- and second-order derivatives of $T^n$, then extend the results to $f(T)$. We show that these derivatives can be obtained as blocks of $f$ evaluated on enlarged block matrices, allowing the sums over open-slot positions (Appendix~\ref{sec:open-slot}) to be computed directly through the same matrix function.

To obtain the first-order derivative, consider the perturbation $T\to T+\epsilon\Delta$. The terms of order $\epsilon$ in $(T+\epsilon\Delta)^n$ are
\begin{equation*}
 \epsilon\sum_{a+b=n-1}T^a\Delta T^b ,
\end{equation*}
with $a,b\geq0$: one of the $n$ factors $T$ becomes an open slot, filled by $\Delta$, and the sum runs over the $n$ positions of the open slot. 

The central mathematical observation in TMC is that this sum can also be expressed as the upper-right block of the $n$th power of a $2\times2$ block matrix,
\begin{equation}
 \left[\begin{pmatrix}T&\Delta\\0&T\end{pmatrix}^n\right]_{12}=\sum_{a+b=n-1}T^a\Delta T^b ,
 \label{eq:block-power}
\end{equation}
where $[\,\cdot\,]_{ij}$ denotes the block in the $i$th block row and the $j$th block column. Equation~\eqref{eq:block-power} follows by induction on $n$.

For the second-order derivative, consider the perturbation $T\to T+\epsilon_1\Delta_1+\epsilon_2\Delta_2$. The terms of order $\epsilon_1\epsilon_2$ in $(T+\epsilon_1\Delta_1+\epsilon_2\Delta_2)^n$ in which $\Delta_1$ stands to the left of $\Delta_2$ are
\begin{equation*}
 \epsilon_1\epsilon_2\sum_{a+b+c=n-2}T^a\Delta_1T^b\Delta_2T^c ,
\end{equation*}
with $a,b,c\geq0$: two of the $n$ factors $T$ become open slots, filled by $\Delta_1$ and $\Delta_2$, and the sum runs over the positions of both open slots. 

The key observation is applied here again: this sum can also be expressed as the upper-right block of the $n$th power of a $3\times3$ block matrix,
\begin{equation}
 \left[\begin{pmatrix}T&\Delta_1&0\\0&T&\Delta_2\\0&0&T\end{pmatrix}^n\right]_{13}=\sum_{a+b+c=n-2}T^a\Delta_1T^b\Delta_2T^c .
 \label{eq:block-power2}
\end{equation}
Equation~\eqref{eq:block-power2} follows by the same induction on $n$. The terms with $\Delta_2$ to the left of $\Delta_1$ are obtained by exchanging $\Delta_1$ and $\Delta_2$.

Since taking a block is linear, multiplying Eqs.~\eqref{eq:block-power} and~\eqref{eq:block-power2} by $c_n$ and summing over $n$ gives, for any $f$,
\begin{align}
 \left[f\!\begin{pmatrix}T&\Delta\\0&T\end{pmatrix}\right]_{12}&=\sum_{n\geq1}c_n\sum_{a+b=n-1}T^a\Delta T^b ,\label{eq:f-block1}\\
 \left[f\!\begin{pmatrix}T&\Delta_1&0\\0&T&\Delta_2\\0&0&T\end{pmatrix}\right]_{13}&=\sum_{n\geq2}c_n\sum_{a+b+c=n-2}T^a\Delta_1T^b\Delta_2T^c .\label{eq:f-block2}
\end{align}
These identities thus provide closed-form expressions for the first- and second-order variations of matrix functions.

The induction used above does not require the diagonal blocks to be equal, so Eqs.~\eqref{eq:block-power}--\eqref{eq:f-block2} also hold when they differ. With $T_1$, $T_2$, and $T_3$ on the diagonal, we write
\begin{equation}
 \begin{aligned}
  f^{[1]}[T_1,T_2](\Delta)&=\left[f\!\begin{pmatrix}T_1&\Delta\\0&T_2\end{pmatrix}\right]_{12},\\
  f^{[2]}[T_1,T_2,T_3](\Delta_1,\Delta_2)&=\left[f\!\begin{pmatrix}T_1&\Delta_1&0\\0&T_2&\Delta_2\\0&0&T_3\end{pmatrix}\right]_{13}.
 \end{aligned}
 \label{eq:fn-notation}
\end{equation}
The superscript of $f$ counts the inserted blocks $\Delta$, with $f^{[0]}=f$; the square brackets list the transfer matrices on the diagonals, and the parentheses hold the inserted blocks. For equal diagonal blocks we write $f^{[1]}[T]$ and $f^{[2]}[T]$: then $f^{[1]}[T](\Delta)$ is the Fr\'echet derivative of $f$ at $T$ in the direction $\Delta$, as in Eq.~\eqref{eq:f-block1}, and $f^{[2]}[T](\Delta,\Delta)$ is half the second derivative of $f(T+\epsilon\Delta)$ at $\epsilon=0$. Left open, the slot is written $\slot$, and $f^{[1]}[T_1,T_2](\slot)$ is the linear map $\Delta\mapsto f^{[1]}[T_1,T_2](\Delta)$ (Appendix~\ref{sec:open-slot}).

\subsection{Energy and gradients}
\label{sec:energy-gradient}
We first summarize the expressions for the iMPS energy and its gradient derived in Ref.~\cite{Yang2026} and introduce the notation used below. Consider pair-interaction terms $J_dX_iY_{i+d}$, summed over sites $i$ and distances $d\geq1$.
For a one-site iMPS tensor $A^s$ of bond dimension $\chi$, define the transfer matrix
\begin{equation}
  T=\sum_s A^s\otimes\overline{A^s}=%
\vcenter{\hbox{\begin{tikzpicture}[x=3.7pt,y=3.7pt]%
 \definecolor{hbl}{HTML}{5271C4}\definecolor{hble}{HTML}{3A55A0}%
 \useasboundingbox (-5.2,-7.6) rectangle (3.0,7.6);%
 \foreach \y in {3.7,-3.7}{\node[font=\scriptsize,inner sep=0.5pt,anchor=east] at (-2.4,\y) {$\chi$};}%
 \foreach \y in {3.7,-3.7}{\draw[line width=1.3pt] (-2.3,\y)--(2.3,\y);}%
 \draw[line width=1.3pt] (0,-2.5)--(0,2.5);%
 \foreach \y in {3.7,-3.7}{\path[fill=hbl,draw=hble,line width=0.5pt] (0,\y) circle (1.2);}%
 \node[font=\scriptsize,inner sep=1pt] at (0,6.3) {$A$};%
 \node[font=\scriptsize,inner sep=1pt] at (0,-6.4) {$\overline{A}$};%
\end{tikzpicture}}}%
.
 \label{eq:transfer-function}
\end{equation}
The bar denotes complex conjugation; $A^s$ forms the ket layer of $T$ and $\overline{A^s}$ its bra layer. More generally, $T(B',B)=\sum_sB^s\otimes\overline{B'^s}$ is the transfer matrix with the tensor $B$ in the ket layer and $B'$ in the bra layer, so $T=T(A,A)$. Inserting an operator $O$ at a site gives $T^O=\sum_{s,t}O_{st}\,A^t\otimes\overline{A^s}$, so $T=T^{\id}$.
We normalize the iMPS so that the dominant eigenvalue of $T$ is $1$. The right and left eigenvectors of $T$ for this eigenvalue, $T\ket{R}=\ket{R}$ and $\bra{L}T=\bra{L}$, normalized as $\braket{L}{R}=1$, are the right and left environments: $\ket{R}$ contracts the half-infinite chain to the right of a site, and $\bra{L}$ the one to its left. The expectation value of each term contains a disconnected part, which comes from the fixed point of $T$. We separate it with the projector $\Qproj$ that removes the fixed point, which defines the connected transfer matrix $\Tconn$:
\begin{equation}
\Qproj=\id-\ket{R}\bra{L},\qquad
\Tconn=\Qproj T\Qproj.
\label{eq:connected-channel}
\end{equation}
Then $T=\ket{R}\bra{L}+\Tconn$ and $\ket{R}\bra{L}\Tconn=\Tconn\ket{R}\bra{L}=0$. The interaction term at distance $d$, $\bra{L}T^XT^{d-1}T^Y\ket{R}$, therefore splits into the product $\langle X\rangle\langle Y\rangle$, which does not decay with $d$ (its sum over $d$ is discussed for the Coulomb chain in Sec.~\ref{sec:coulomb}), and a connected part in which only $\Tconn$ appears, between $\bra{L_X}=\bra{L}T^X\Qproj$ and $\ket{R_Y}=\Qproj T^Y\ket{R}$. The connected interaction energy per site is
\begin{equation}
 \begin{split}
e_J^{\mathrm c}
&=\sum_{d\geq1}J_d\bra{L_X}\Tconn^{d-1}\ket{R_Y}\\
&=\sum_{d\geq1}J_d\,%
\vcenter{\hbox{\begin{tikzpicture}[x=3.7pt,y=3.7pt,
  hline/.style={line width=1.3pt},
  hop/.style={draw=hgre,fill=hgr,rounded corners=1.5pt,minimum size=10.4pt,inner sep=0pt,line width=0.5pt,font=\scriptsize},
  henv/.style={draw=hgre,fill=hgr,rounded corners=1.5pt,line width=0.5pt},
  hq/.style={draw=black!55,fill=black!8,rounded corners=1.5pt,line width=0.5pt}]%
 \definecolor{hbl}{HTML}{5271C4}\definecolor{hble}{HTML}{3A55A0}%
 \definecolor{hgr}{HTML}{D6EAC6}\definecolor{hgre}{HTML}{7FA85F}%
 \def\ecol#1{\draw[hline] (#1,-2.5)--(#1,2.5);\foreach \y in {3.7,-3.7}{\path[fill=hbl,draw=hble,line width=0.5pt] (#1,\y) circle (1.2);}}%
 \def\eop#1#2{\ecol{#1}\node[hop] at (#1,0) {$#2$};}%
 \def\ebox#1#2#3#4{\path[#1] ({#2-#3},-4.4) rectangle ({#2+#3},4.4);\node[font=\scriptsize,inner sep=0pt] at (#2,0) {$#4$};}%
 \def\eparl#1{\draw[line width=0.7pt] ({#1+0.9},5.6) .. controls ({#1-0.54},2.52) and ({#1-0.54},-2.52) .. ({#1+0.9},-5.6);}%
 \def\eparr#1{\draw[line width=0.7pt] ({#1-0.9},5.6) .. controls ({#1+0.54},2.52) and ({#1+0.54},-2.52) .. ({#1-0.9},-5.6);}%
 \useasboundingbox (-2.4,-6.4) rectangle (32.7,7.6);%
 \foreach \y in {3.7,-3.7}{\draw[hline] (0,\y)--(30.3,\y);}%
 \ebox{henv}{0}{2}{L}%
 \eop{5.6}{X}%
 \ebox{hq}{9.8}{1.3}{P}%
 \eparl{12.4}%
 \ecol{15.2}%
 \eparr{18.0}%
 \node[font=\scriptsize,anchor=west,inner sep=0pt] at (18.1,6.5) {$d-1$};%
 \ebox{hq}{21.3}{1.3}{P}%
 \eop{25.3}{Y}%
 \ebox{henv}{30.3}{2}{R}%
\end{tikzpicture}}}%
\\
&=\bra{L_X}F_J(\Tconn)\ket{R_Y}\\
&=%
\vcenter{\hbox{\begin{tikzpicture}[x=3.7pt,y=3.7pt,
  hline/.style={line width=1.3pt},
  hop/.style={draw=hgre,fill=hgr,rounded corners=1.5pt,minimum size=10.4pt,inner sep=0pt,line width=0.5pt,font=\scriptsize},
  henv/.style={draw=hgre,fill=hgr,rounded corners=1.5pt,line width=0.5pt},
  hq/.style={draw=black!55,fill=black!8,rounded corners=1.5pt,line width=0.5pt}]%
 \definecolor{hbl}{HTML}{5271C4}\definecolor{hble}{HTML}{3A55A0}%
 \definecolor{hgr}{HTML}{D6EAC6}\definecolor{hgre}{HTML}{7FA85F}%
 \def\ecol#1{\draw[hline] (#1,-2.5)--(#1,2.5);\foreach \y in {3.7,-3.7}{\path[fill=hbl,draw=hble,line width=0.5pt] (#1,\y) circle (1.2);}}%
 \def\eop#1#2{\ecol{#1}\node[hop] at (#1,0) {$#2$};}%
 \def\ebox#1#2#3#4{\path[#1] ({#2-#3},-4.4) rectangle ({#2+#3},4.4);\node[font=\scriptsize,inner sep=0pt] at (#2,0) {$#4$};}%
 \def\eparl#1{\draw[line width=0.7pt] ({#1+0.9},5.6) .. controls ({#1-0.54},2.52) and ({#1-0.54},-2.52) .. ({#1+0.9},-5.6);}%
 \def\eparr#1{\draw[line width=0.7pt] ({#1-0.9},5.6) .. controls ({#1+0.54},2.52) and ({#1+0.54},-2.52) .. ({#1-0.9},-5.6);}%
 \useasboundingbox (-2.4,-6.4) rectangle (36.5,6.4);%
 \foreach \y in {3.7,-3.7}{\draw[hline] (0,\y)--(34.1,\y);}%
 \ebox{henv}{0}{2}{L}%
 \eop{5.6}{X}%
 \ebox{hq}{9.8}{1.3}{P}%
 \node[font=\small,inner sep=0pt,fill=white] at (13.9,0) {$F_J$};%
 \eparl{16.2}%
 \ecol{19.0}%
 \eparr{21.8}%
 \ebox{hq}{25.1}{1.3}{P}%
 \eop{29.1}{Y}%
 \ebox{henv}{34.1}{2}{R}%
\end{tikzpicture}}}%
.
\end{split}
 \label{eq:energy}
\end{equation}
Here $F_J(z)=\sum_{d\geq1}J_dz^{d-1}$ is the generating function of the interaction, in which the power $d-1$ counts the transfer matrices between the two sites of a term at distance $d$. Since $\Qproj$ commutes with $T$ and $\Qproj^2=\Qproj$, only the projectors $\Qproj$ adjacent to $X$ and $Y$ remain. The projectors are necessary: without them, $F_J(T)$ would contain $\ket{R}\bra{L}$ with weight $F_J(1)=\sum_dJ_d$, which diverges for $J_d=V/d^\alpha$ with $\alpha\le1$~\footnote{In the matrix-free evaluation (Appendix~\ref{sec:matrix-free}) we therefore apply $\protect\Tconn$ at every step, because rounding errors would otherwise build up a component along $\ket{R}$.}. In this formalism, the interaction coefficients $J_d$ enter the energy only through their generating function $F_J$, and the state only through $T$.

The series in Eq.~\eqref{eq:energy} converges if the spectrum of $\Tconn$ lies inside the disk of convergence of $F_J(z)$. For a power-law interaction this is the unit disk, so it suffices that the fixed-point eigenvalue $1$ of $T$ is nondegenerate and all other eigenvalues have modulus below one. For $J_d=V/d^\alpha$ one has $F_J(z)=V\Li_\alpha(z)/z$~\cite{Yang2026}, where $\Li_\alpha(z)/z=\sum_{d\geq1}z^{d-1}/d^\alpha$ has constant term $1$. Hence, $F_J(0)=J_1=V$ and $z=0$ is a removable singularity. For the Coulomb case, $F_J(z)=-V\ln(1-z)/z$. For a power law, $F_J$ is transcendental, whereas a finite MPO gives a rational $F_J$ (Appendix~\ref{sec:mpo}). At $z=1$, where $F_J$ diverges for $\alpha\le1$, $\Qproj$ keeps the eigenvalue $1$ out of $\Tconn$, so the energy is finite~\footnote{An analogous formalism exists for continuous MPS (cMPS)~\cite{Haegeman2013cMPS}, where the sum over $d$ becomes an integral and $F_J$ becomes the Laplace transform of the interaction.}.

A natural way to optimize an iMPS variationally is gradient-based minimization of its energy~\cite{Xie2020,Hauru2021}. In TMC the connected part of the interaction energy is the single contraction $e_J^{\mathrm c}=\bra{L_X}F_J(\Tconn)\ket{R_Y}$ of Eq.~\eqref{eq:energy}, in which the MPS tensor enters through the two ends $\bra{L_X}$ and $\ket{R_Y}$ and through the string $\Tconn$ between them. The gradient with respect to $\overline{A}$ is the coefficient of $\overline{\delta A}$ in the first-order perturbation of $e_J^{\mathrm c}$. A perturbation $\delta A$ changes the transfer matrix by $\delta T=\sum_s(\delta A^s\otimes\overline{A^s}+A^s\otimes\overline{\delta A^s})$ and $\Tconn$ by $\delta\Tconn$, and by Eq.~\eqref{eq:f-block1}
\begin{equation}
 \begin{split}
 \delta e_J^{\mathrm c}={}&\bra{\delta L_X}F_J(\Tconn)\ket{R_Y}+\bra{L_X}F_J(\Tconn)\ket{\delta R_Y}\\
 &+\bra{L_X}\left[F_J\!\begin{pmatrix}\Tconn&\delta\Tconn\\0&\Tconn\end{pmatrix}\right]_{12}\ket{R_Y}.
 \end{split}
 \label{eq:gradient}
\end{equation}
The first two terms come from the endpoints and are evaluated with automatic differentiation through the dominant eigensolver~\cite{Xie2020}. The last term is the open-slot sum of Sec.~\ref{sec:open-sums} with $\delta\Tconn$ in the slot, $\bra{L_X}F_J^{[1]}[\Tconn](\delta\Tconn)\ket{R_Y}$, in the notation of Eq.~\eqref{eq:fn-notation} for $f=F_J$. Written out,
\begin{equation}
 F_J^{[1]}[T_1,T_2](\Delta)=\sum_{d\geq2}J_d\sum_{a+b=d-2}T_1^a\Delta T_2^b ,
 \label{eq:FJ1}
\end{equation}
and $F_J^{[1]}[T](\Delta)$ stands for $T_1=T_2=T$. For the gradient, the derivative with respect to $\overline{A}$ removes the bra tensor from the contraction and leaves an open slot (Appendix~\ref{sec:open-slot}), and $\delta T$ becomes $T(\slot,A)$. Appendix~\ref{sec:matrix-free} describes how Eq.~\eqref{eq:gradient} is evaluated efficiently with Krylov methods, and Ref.~\cite{Yang2026} describes its use in variational optimization with L-BFGS.

\subsection{Effective Hamiltonians for iDMRG and VUMPS}
\label{sec:effective-operators}
The iDMRG~\cite{White1992,McCulloch2008} and VUMPS~\cite{Zauner2018,Vanderstraeten2019} algorithms are well-established methods for ground-state calculations in one-dimensional quantum systems~\cite{Schollwock2011}. While iDMRG is based on successive growth of the chain with truncation of the bond dimension and VUMPS on direct variational optimization, both reduce the global many-body problem to local eigenvalue problems for effective Hamiltonians by compressing the environment into finite-dimensional Hilbert spaces. Therefore, the key to performing iDMRG and VUMPS calculations lies in finding a representation of the effective Hamiltonian.

Although traditional methods typically build the effective Hamiltonian from an MPO, Ref.~\cite{Haegeman2013Tangent} provides an alternative definition—one that is more useful for understanding the approach presented in this work: the Hessian operator obtained by restricting the global energy functional to a local MPS variational space.

\newdimen\hu \hu=3.7pt
\definecolor{hbl}{HTML}{5271C4}\definecolor{hble}{HTML}{3A55A0}
\definecolor{hgr}{HTML}{D6EAC6}\definecolor{hgre}{HTML}{7FA85F}
\definecolor{hor}{HTML}{F6C58B}\definecolor{hore}{HTML}{C77A1E}
\definecolor{hBfill}{HTML}{C2BFD7}\definecolor{hBedge}{HTML}{827B9D}
\tikzset{
  hline/.style={line width=1.3pt},
  hdot/.style={line width=1.3pt,line cap=round,dash pattern=on 0pt off 2.5pt},
  hop/.style={draw=hgre,fill=hgr,rounded corners=1.5pt,minimum size=2.8\hu,inner sep=0pt,line width=0.5pt,font=\scriptsize},
  hopw/.style={hop,draw=hore,fill=hor},
}
\newcommand{\hten}[3][1.2]{\path[fill=hbl,draw=hble,line width=0.5pt] (#2,#3) circle (#1);}
\newcommand{\hhole}[3][1.2]{\path[fill=white,draw=hble,line width=0.8pt,dash pattern=on 1.6pt off 1.3pt] (#2,#3) circle (#1);}
\newcommand{\hcol}[2]{\draw[hline] (#1,{#2-2.5})--(#1,{#2+2.5});\hten{#1}{#2+3.7}\hten{#1}{#2-3.7}}
\newcommand{\hcolh}[2]{\draw[hline] (#1,{#2-2.5})--(#1,{#2+2.5});\hhole{#1}{#2+3.7}\hhole{#1}{#2-3.7}}
\newcommand{\hcolop}[3]{\hcol{#1}{#2}\node[hop] at (#1,#2) {$#3$};}
\newcommand{\hcolc}[2]{\draw[hline] (#1,{#2-1.05})--(#1,{#2+1.05});\hten[0.95]{#1}{#2+2}\hten[0.95]{#1}{#2-2}}
\newcommand{\hcolch}[2]{\draw[hline] (#1,{#2-1.05})--(#1,{#2+1.05});\hhole[0.95]{#1}{#2+2}\hhole[0.95]{#1}{#2-2}}
\newcommand{\hparl}[4]{\draw[line width=0.7pt] (#1+#4,{#2+#3}) .. controls (#1-0.6*#4,{#2+0.45*#3}) and (#1-0.6*#4,{#2-0.45*#3}) .. (#1+#4,{#2-#3});}
\newcommand{\hparr}[4]{\draw[line width=0.7pt] (#1-#4,{#2+#3}) .. controls (#1+0.6*#4,{#2+0.45*#3}) and (#1+0.6*#4,{#2-0.45*#3}) .. (#1-#4,{#2-#3});}
\newcommand{\hends}[2]{\foreach \y in {#1}{\draw[hline] (-2.3,\y)--(0,\y);\draw[hline] (#2,\y)--(#2+2.3,\y);\draw[hdot] (-5,\y)--(-3.2,\y);\draw[hdot] (#2+3.2,\y)--(#2+5,\y);}}

\newcommand{\hnetblob}{\vcenter{\hbox{\begin{tikzpicture}[x=\hu,y=\hu]
 \useasboundingbox (-5.2,-6.5) rectangle (27.6,6.5);
 \foreach \g/\o in {2.4/0.07,1.6/0.10,0.8/0.14,0/0.22}{\fill[hore,opacity=\o,rounded corners=4pt] ({-3.2-\g},{-1.9-0.35*\g}) rectangle ({25.6+\g},{1.9+0.35*\g});}
 \draw[hline] (0,3.7)--(22.4,3.7); \draw[hline] (0,-3.7)--(22.4,-3.7);
 \foreach \y in {3.7,-3.7}{\draw[hline] (-2.3,\y)--(0,\y);\draw[hline] (22.4,\y)--(24.7,\y);\draw[hdot] (-5,\y)--(-3.2,\y);\draw[hdot] (25.5,\y)--(27.4,\y);}
 \foreach \x in {0,5.6,11.2,16.8,22.4}{\draw[hline] (\x,1.9)--(\x,3.7);\draw[hline] (\x,-1.9)--(\x,-3.7);}
 \foreach \x in {0,5.6,16.8,22.4}{\hten{\x}{3.7}\hten{\x}{-3.7}}
 \hhole{11.2}{3.7}\hhole{11.2}{-3.7}
 \node[font=\scriptsize] at (11.2,-6.3) {$i$};
 \node[font=\small,text=black] at (11.2,0) {$H$}; %
\end{tikzpicture}}}}

\newcommand{\hnetmpo}{\vcenter{\hbox{\begin{tikzpicture}[x=\hu,y=\hu]
 \useasboundingbox (-5.2,-6.5) rectangle (27.6,6.5);
 \draw[hline] (0,3.7)--(22.4,3.7); \draw[hline] (0,-3.7)--(22.4,-3.7); \draw[hline] (-2.3,0)--(24.7,0);
 \foreach \y in {3.7,-3.7}{\draw[hline] (-2.3,\y)--(0,\y);\draw[hline] (22.4,\y)--(24.7,\y);}
 \foreach \y in {3.7,0,-3.7}{\draw[hdot] (-5,\y)--(-3.2,\y);\draw[hdot] (25.5,\y)--(27.4,\y);}
 \foreach \x in {0,5.6,11.2,16.8,22.4}{\draw[hline] (\x,-3.7)--(\x,3.7);}
 \foreach \x in {0,5.6,16.8,22.4}{\hten{\x}{3.7}\hten{\x}{-3.7}}
 \hhole{11.2}{3.7}\hhole{11.2}{-3.7}
 \node[font=\scriptsize] at (11.2,-6.3) {$i$};
 \foreach \x in {0,5.6,11.2,16.8,22.4}{\node[hopw] at (\x,0) {$W$};}
\end{tikzpicture}}}}

\newcommand{\hnetsum}{\vcenter{\hbox{\begin{tikzpicture}[x=\hu,y=\hu]
 \useasboundingbox (-5.2,-6.5) rectangle (27.6,6.5);
 \draw[hline] (0,3.7)--(22.4,3.7); \draw[hline] (0,-3.7)--(22.4,-3.7);
 \hends{3.7,-3.7}{22.4}
 \hcolop{0}{0}{X}\hcol{5.6}{0}\hcolh{11.2}{0}\hcol{16.8}{0}\hcolop{22.4}{0}{Y}
 \hparl{2.9}{0}{5.6}{0.9}\hparr{8.3}{0}{5.6}{0.9}\hparl{14.1}{0}{5.6}{0.9}\hparr{19.5}{0}{5.6}{0.9}
 \node[font=\scriptsize] at (9.1,5.9) {$a$}; \node[font=\scriptsize] at (20.3,5.9) {$b$};
 \node[font=\scriptsize] at (11.2,-6.3) {$i$};
\end{tikzpicture}}}}

\newcommand{\hbrl}[3]{\draw[line width=0.8pt] (#1+1.2,{#2+#3})--(#1,{#2+#3})--(#1,{#2-#3})--(#1+1.2,{#2-#3});}
\newcommand{\hbrr}[3]{\draw[line width=0.8pt] (#1-1.2,{#2+#3})--(#1,{#2+#3})--(#1,{#2-#3})--(#1-1.2,{#2-#3});}
\newcommand{\hnetF}{\vcenter{\hbox{\begin{tikzpicture}[x=1.1\hu,y=1.1\hu]
 \useasboundingbox (-5.2,-9.8) rectangle (35.3,9.8);
 \foreach \y in {3.7,-3.7}{\draw[hline] (-2.3,\y)--(0,\y);\draw[hline] (30.1,\y)--(32.4,\y);\draw[hdot] (-5,\y)--(-3.2,\y);\draw[hdot] (33.3,\y)--(35.1,\y);
   \draw[hline] (0,\y)--(2.6,\y);\draw[hline] (26.7,\y)--(30.1,\y);}   %
 \hcolop{0}{0}{X}\hcolop{30.1}{0}{Y}
 \hbrl{2.6}{0}{8.6}\hbrr{26.7}{0}{8.6}
 \node[font=\scriptsize] at (28.2,-9.3) {$12$};
 \node at (6.4,0) {\Large $F_J$};
 \hparl{9.6}{0}{7.7}{1.9}\hparr{22.9}{0}{7.7}{1.9}
 \foreach \y in {2,-2}{\draw[hline] (11.9,{3.9+\y})--(15.1,{3.9+\y}); \draw[hline] (17.1,{3.9+\y})--(20.3,{3.9+\y});
   \draw[hline] (17.1,{-3.9+\y})--(20.3,{-3.9+\y});}
 \hcolc{13.5}{3.9}\hcolch{18.7}{3.9}\hcolc{18.7}{-3.9}
 \node at (13.5,-3.9) {\large $0$};
\end{tikzpicture}}}}

\newcommand{\hcolbh}[2]{\draw[hline] (#1,{#2-2.5})--(#1,{#2+2.5});\hten{#1}{#2+3.7}\hhole{#1}{#2-3.7}}
\newcommand{\hcolkh}[2]{\draw[hline] (#1,{#2-2.5})--(#1,{#2+2.5});\hhole{#1}{#2+3.7}\hten{#1}{#2-3.7}}
\newcommand{\hcolcbh}[2]{\draw[hline] (#1,{#2-1.05})--(#1,{#2+1.05});\hten[0.95]{#1}{#2+2}\hhole[0.95]{#1}{#2-2}}
\newcommand{\hcolckh}[2]{\draw[hline] (#1,{#2-1.05})--(#1,{#2+1.05});\hhole[0.95]{#1}{#2+2}\hten[0.95]{#1}{#2-2}}
\newcommand{\hstub}[3]{\foreach \s in {2,-2}{\draw[hline] ({#1-1.6},{#2+\s})--({#1+1.6},{#2+\s});}#3}
\newcommand{\hlab}[3]{\node[font=\tiny,anchor=west,inner sep=1pt] at ({#1+0.2},#2) {$#3$};}
\newcommand{\hbrace}[4]{\draw[line width=0.6pt,rounded corners=1.3pt] (#1,#4)--(#1,{#4-0.7})--({#2-0.9},{#4-0.7})--(#2,{#4-1.5})--({#2+0.9},{#4-0.7})--(#3,{#4-0.7})--(#3,#4);}

\newcommand{\hBten}[3]{\path[fill=hBfill,draw=hBedge,line width=0.5pt] (#1,#2) circle (1.2);
  \node[font=\scriptsize,inner sep=0pt] at (#1,#2) {$#3$};}
\newcommand{\hnetansatz}{\vcenter{\hbox{\begin{tikzpicture}[x=\hu,y=\hu]
 \useasboundingbox (-5.2,-3.6) rectangle (27.6,4.2);
 \draw[hline] (-2.3,0)--(24.7,0);
 \draw[hdot] (-5,0)--(-3.2,0);\draw[hdot] (25.5,0)--(27.4,0);
 \foreach \x in {0,5.6,11.2,16.8,22.4}{\draw[hline] (\x,0)--(\x,-3.0);}
 \foreach \x in {0,5.6,16.8,22.4}{\hten{\x}{0}}
 \hBten{11.2}{0}{B}
 \node[font=\scriptsize] at (11.2,3.1) {$j$};
\end{tikzpicture}}}}
\newcommand{\hnetexcpair}{\vcenter{\hbox{\begin{tikzpicture}[x=\hu,y=\hu]
 \useasboundingbox (-5.2,-7.9) rectangle (27.6,7.9);
 \foreach \y in {3.7,0,-3.7}{\draw[hline] (-2.3,\y)--(24.7,\y);
   \draw[hdot] (-5,\y)--(-3.2,\y);\draw[hdot] (25.5,\y)--(27.4,\y);}
 \foreach \x in {0,5.6,11.2,16.8,22.4}{\draw[hline] (\x,-3.7)--(\x,3.7);}
 \foreach \x in {0,11.2,22.4}{\hten{\x}{3.7}\hten{\x}{-3.7}}
 \hten{5.6}{3.7}\hBten{5.6}{-3.7}{B'}
 \hBten{16.8}{3.7}{B}\hten{16.8}{-3.7}
 \node[font=\scriptsize] at (5.6,-6.7) {$0$};
 \node[font=\scriptsize] at (16.8,6.7) {$d$};
 \foreach \x in {0,5.6,11.2,16.8,22.4}{\node[hopw] at (\x,0) {$W$};}
\end{tikzpicture}}}}
\newcommand{\hcolkB}[2]{\draw[hline] (#1,{#2-2.5})--(#1,{#2+2.5});
  \hBten{#1}{#2+3.7}{B}\hten{#1}{#2-3.7}}
\newcommand{\hcolckB}[2]{\draw[hline] (#1,{#2-1.05})--(#1,{#2+1.05});
  \path[fill=hBfill,draw=hBedge,line width=0.5pt] (#1,{#2+2}) circle (0.95);
  \node[font=\tiny,inner sep=0pt] at (#1,{#2+2}) {$B$};\hten[0.95]{#1}{#2-2}}

\newcommand{\hnetexcblob}{\vcenter{\hbox{\begin{tikzpicture}[x=\hu,y=\hu]
 \useasboundingbox (-5.2,-7.9) rectangle (27.6,7.9);
 \foreach \g/\o in {2.4/0.07,1.6/0.10,0.8/0.14,0/0.22}{\fill[hore,opacity=\o,rounded corners=4pt] ({-3.2-\g},{-1.9-0.35*\g}) rectangle ({25.6+\g},{1.9+0.35*\g});}
 \draw[hline] (0,3.7)--(22.4,3.7); \draw[hline] (0,-3.7)--(22.4,-3.7);
 \foreach \y in {3.7,-3.7}{\draw[hline] (-2.3,\y)--(0,\y);\draw[hline] (22.4,\y)--(24.7,\y);\draw[hdot] (-5,\y)--(-3.2,\y);\draw[hdot] (25.5,\y)--(27.4,\y);}
 \foreach \x in {0,5.6,11.2,16.8,22.4}{\draw[hline] (\x,1.9)--(\x,3.7);\draw[hline] (\x,-1.9)--(\x,-3.7);}
 \foreach \x in {0,11.2,22.4}{\hten{\x}{3.7}\hten{\x}{-3.7}}
 \hten{5.6}{3.7}\hhole{5.6}{-3.7}\hBten{16.8}{3.7}{B}\hten{16.8}{-3.7}
 \node[font=\scriptsize] at (5.6,-6.7) {$0$};\node[font=\scriptsize] at (16.8,6.7) {$d$};
 \node[font=\small,text=black] at (11.2,0) {$H$};
\end{tikzpicture}}}}
\newcommand{\hnetexcmpo}{\vcenter{\hbox{\begin{tikzpicture}[x=\hu,y=\hu]
 \useasboundingbox (-5.2,-7.9) rectangle (27.6,7.9);
 \draw[hline] (0,3.7)--(22.4,3.7); \draw[hline] (0,-3.7)--(22.4,-3.7); \draw[hline] (-2.3,0)--(24.7,0);
 \foreach \y in {3.7,-3.7}{\draw[hline] (-2.3,\y)--(0,\y);\draw[hline] (22.4,\y)--(24.7,\y);}
 \foreach \y in {3.7,0,-3.7}{\draw[hdot] (-5,\y)--(-3.2,\y);\draw[hdot] (25.5,\y)--(27.4,\y);}
 \foreach \x in {0,5.6,11.2,16.8,22.4}{\draw[hline] (\x,-3.7)--(\x,3.7);}
 \foreach \x in {0,11.2,22.4}{\hten{\x}{3.7}\hten{\x}{-3.7}}
 \hten{5.6}{3.7}\hhole{5.6}{-3.7}\hBten{16.8}{3.7}{B}\hten{16.8}{-3.7}
 \node[font=\scriptsize] at (5.6,-6.7) {$0$};\node[font=\scriptsize] at (16.8,6.7) {$d$};
 \foreach \x in {0,5.6,11.2,16.8,22.4}{\node[hopw] at (\x,0) {$W$};}
\end{tikzpicture}}}}
\newcommand{\hnetexcsum}{\vcenter{\hbox{\begin{tikzpicture}[x=1.0\hu,y=1.0\hu]
 \useasboundingbox (-5.2,-6.5) rectangle (34,6.5);
 \draw[hline] (0,3.7)--(28.8,3.7); \draw[hline] (0,-3.7)--(28.8,-3.7);
 \hends{3.7,-3.7}{28.8}
 \hcolop{0}{0}{X}\hcol{4.8}{0}\hcolbh{9.6}{0}\hcol{14.4}{0}\hcolkB{19.2}{0}\hcol{24}{0}\hcolop{28.8}{0}{Y}
 \foreach \x/\l in {4.8/a,14.4/b,24/c}{\hparl{\x-2.4}{0}{5.6}{0.9}\hparr{\x+2.4}{0}{5.6}{0.9}
   \node[font=\scriptsize] at (\x+3.1,6) {$\l$};}
\end{tikzpicture}}}}
\newcommand{\hnetexcF}{\vcenter{\hbox{\begin{tikzpicture}[x=\hu,y=\hu]
 \useasboundingbox (-5.2,-15.4) rectangle (46.5,15.4);
 \foreach \y in {3.7,-3.7}{\draw[hline] (0,\y)--(2.6,\y);\draw[hline] (37.8,\y)--(41.2,\y);
   \draw[hline] (-2.3,\y)--(0,\y);\draw[hline] (41.2,\y)--(43.5,\y);
   \draw[hdot] (-5,\y)--(-3.2,\y);\draw[hdot] (44.3,\y)--(46.3,\y);}
 \hcolop{0}{0}{X}\hcolop{41.2}{0}{Y}
 \hbrl{2.6}{0}{14.2}\hbrr{37.8}{0}{14.2}
 \node[font=\scriptsize] at (39.4,-14.7) {$13$};
 \node at (6.0,0) {\large $F_J$};
 \hparl{9.4}{0}{13.4}{1.9}\hparr{35.0}{0}{13.4}{1.9}
 \hstub{13.7}{9.4}{\hcolc{13.7}{9.4}}
 \hstub{22.5}{9.4}{\hcolcbh{22.5}{9.4}}
 \hstub{22.5}{0}{\hcolc{22.5}{0}}
 \node[font=\scriptsize,anchor=east,inner sep=0pt] at (20.2,0) {$\e^{\ii k}$};
 \hstub{31.7}{0}{\hcolckB{31.7}{0}}
 \node[font=\scriptsize,anchor=east,inner sep=0pt] at (29.4,0) {$\e^{\ii k}$};
 \hstub{31.7}{-9.4}{\hcolc{31.7}{-9.4}}
 \foreach \p in {(31.7,9.4),(13.7,0),(13.7,-9.4),(22.5,-9.4)}{\node at \p {\large $0$};}
\end{tikzpicture}}}}

\newcommand{\hnetfqstate}{\hbox{\begin{tikzpicture}[baseline=-1.1ex,x=\hu,y=\hu]
 \useasboundingbox (-4.0,-5.6) rectangle (16.8,2.2);
 \draw[hline] (-1.9,0)--(14.7,0);
 \draw[hdot] (-3.8,0)--(-2.7,0);\draw[hdot] (15.5,0)--(16.6,0);
 \foreach \x/\lab in {0/{q_{i-1}},6.4/{q_i},12.8/{q_{i+1}}}{
   \draw[hline] (\x,0)--(\x,-3.2);
   \hten[1.6]{\x}{0}
   \node[font=\small,text=white,inner sep=0pt] at (\x,0) {$A$};
   \node[font=\scriptsize,anchor=north,inner sep=1pt] at (\x,-3.2) {$\lab$};}
\end{tikzpicture}}}

\newcommand{\hnetfqLerch}{\vcenter{\hbox{\begin{tikzpicture}[x=\hu,y=\hu]
 \useasboundingbox (-2.5,-7.6) rectangle (32.5,7.6);
 \draw[draw=hgre,fill=hgr,rounded corners=1.5pt,line width=0.5pt] (-2,-4.4) rectangle (2,4.4);
 \node[font=\scriptsize] at (0,0) {$L_a$};
 \draw[draw=hgre,fill=hgr,rounded corners=1.5pt,line width=0.5pt] (28,-4.4) rectangle (32,4.4);
 \node[font=\scriptsize] at (30,0) {$R_b$};
 \foreach \y in {3.7,-3.7}{\draw[hline] (2,\y)--(3.8,\y);\draw[hline] (26.2,\y)--(28,\y);
   \draw[hline] (13.2,\y)--(22.0,\y);}
 \hbrl{3.8}{0}{6.6}\hbrr{26.2}{0}{6.6}
 \node[font=\small] at (7.8,0) {$F_{\alpha,ab}$};
 \hparl{11.5}{0}{5.8}{0.9}\hparr{23.5}{0}{5.8}{0.9}
 \hcol{14.8}{0}\hcol{20.4}{0}
\end{tikzpicture}}}}
\newcommand{\hnetfqkin}{\vcenter{\hbox{\begin{tikzpicture}[x=\hu,y=\hu]
 \useasboundingbox (-4.6,-5.6) rectangle (14.2,5.6);
 \draw[draw=hgre,fill=hgr,rounded corners=1.5pt,line width=0.5pt] (-4.4,-4.6) rectangle (-1.6,4.6);
 \node[font=\scriptsize] at (-3.0,0) {$L$};
 \draw[draw=hgre,fill=hgr,rounded corners=1.5pt,line width=0.5pt] (11.2,-4.6) rectangle (14.0,4.6);
 \node[font=\scriptsize] at (12.6,0) {$R$};
 \foreach \y in {3.7,-3.7}{\draw[hline] (-1.6,\y)--(11.2,\y);}
 \foreach \x in {2.0,7.6}{\draw[hline] (\x,-3.7)--(\x,3.7);\hten{\x}{3.7}\hten{\x}{-3.7}}
 \draw[draw=hore,fill=hor,rounded corners=1.5pt,line width=0.5pt] (0.6,-1.3) rectangle (9.0,1.3);
 \node[font=\small] at (4.8,0) {$h$};
\end{tikzpicture}}}}
\newcommand{\hfqlattice}[2]{%
 \draw[hline] (-2.0,#1)--(32.8,#1);
 \foreach \s in {0,...,11}{\fill[white,draw=hble,line width=0.5pt] ({2.8*\s},#1) circle (0.6);}
 \foreach \s in {#2}{\hten[1.1]{2.8*\s}{#1}}}
\newcommand{\hfqgapbrace}[4]{\draw[line width=0.5pt] (#1,{#3-1.7})--(#1,{#3-2.3})--(#2,{#3-2.3})--(#2,{#3-1.7});
 \node[font=\scriptsize] at ({(#1+#2)/2},{#3-3.5}) {$#4$};}
\newcommand{\hnetfqhopii}{\vcenter{\hbox{\begin{tikzpicture}[x=\hu,y=\hu]
 \useasboundingbox (-2.4,-16.2) rectangle (33.2,5.2);
 \hfqgapbrace{2.8}{11.2}{0}{q}\hfqgapbrace{11.2}{25.2}{0}{p}
 \hfqgapbrace{2.8}{14.0}{-10.5}{q+1}\hfqgapbrace{14.0}{25.2}{-10.5}{p-1}
 \draw[->,line width=0.9pt,>=stealth,hble] (11.2,1.6) to[bend left=70] (14.0,1.6);
 \node[font=\scriptsize] at (12.6,4.5) {$-1$};
 \hfqlattice{0}{1,4,9}
 \hfqlattice{-10.5}{1,5,9}
\end{tikzpicture}}}}
\newcommand{\hnetfqdisp}{\vcenter{\hbox{\begin{tikzpicture}[x=\hu,y=\hu]
 \useasboundingbox (-2.4,-5.4) rectangle (35.6,10.0);
 \foreach \n/\x in {0/2.8,1/8.4,2/14.0,3/19.6,4/25.2}{\draw[gray,line width=0.6pt] (\x,3.6)--(\x,5.0);}
 \draw[gray,line width=0.4pt,dashed] (-2.0,4.3)--(30.8,4.3);
 \node[font=\scriptsize,text=gray,anchor=west] at (31.0,4.3) {$\mu i$};
 \draw[line width=0.5pt] (2.8,6.4)--(2.8,7.0)--(19.6,7.0)--(19.6,6.4);
 \node[font=\scriptsize] at (11.2,8.6) {$\mu s$};
 \draw[->,>=stealth,line width=0.8pt,hore] (2.8,3.4)--(5.6,1.6);
 \node[font=\scriptsize,text=hore,anchor=west] at (5.0,2.6) {$\delta_i$};
 \draw[->,>=stealth,line width=0.8pt,hore] (19.6,3.4)--(16.8,1.6);
 \node[font=\scriptsize,text=hore,anchor=east] at (17.4,2.6) {$\delta_{i+s}$};
 \draw[hline] (-2.0,0)--(30.8,0);
 \foreach \s in {0,...,10}{\fill[white,draw=hble,line width=0.5pt] ({2.8*\s},0) circle (0.6);}
 \foreach \s in {2,4,5,6,9}{\hten[1.1]{2.8*\s}{0}}
 \hfqgapbrace{5.6}{16.8}{0}{d}
\end{tikzpicture}}}}
We first discuss the effective Hamiltonian on a single site, which is required by the VUMPS method. The effective Hamiltonian at the center site $i$ is the second derivative of the energy with respect to the ket tensor $A_i$ and the bra tensor $\overline{A_i}$ there, $H_{\rm eff}=\partial^2E/\partial\overline{A_i}\,\partial A_i$. In diagrammatic notation, differentiating removes the two tensors at site $i$ (dashed circles) from the Hamiltonian contraction, as in automatic differentiation (Appendix~\ref{sec:open-slot}). The open slot left in their place has uncontracted bra and ket indices, and the remaining contraction below is $H_{\rm eff}$, an operator on the center tensor.
{\allowdisplaybreaks
\begin{flalign}
&H_{\rm eff}=\frac{\partial^2E}{\partial\overline{A_i}\,\partial A_i}=\,\hnetblob&&\notag\\[3pt]
\intertext{\raggedright In the MPO representation, the Hamiltonian $H$ is replaced by a chain of MPO tensors $W$~\cite{McCulloch2008,Schollwock2011}:}
&{}=\frac{\partial^2E}{\partial\overline{A_i}\,\partial A_i}=\,\hnetmpo&&\notag\\[3pt]
\intertext{\raggedright Expanding the Hamiltonian into its terms instead, we first consider a term whose ends $X$ and $Y$ lie on opposite sides of the open slot. Let $a$ be the number of transfer matrices between $X$ and the open slot, and $b$ the number between the open slot and $Y$; the term then joins two sites at distance $a+b+2$ and has weight $J_{a+b+2}$:}
&H_{\rm eff} =\sum_{a,b\ge0}J_{a+b+2}\,\hnetsum+\cdots&&\notag\\[3pt]
\intertext{\raggedright where $+\cdots$ collects the remaining terms: those with the open slot on $X$ or $Y$ or outside the interaction string, and the disconnected parts. The sum over $a$ and $b$ is $F_J^{[1]}$ of Eq.~\eqref{eq:FJ1}, the upper-right block of $F_J$ applied to a block-triangular matrix:}
&{}=\hnetF+\cdots&&\notag\\[3pt]
&{}=\bra{L_X}F_J^{[1]}[\Tconn]\bigl(T(\slot,\slot)\bigr)\ket{R_Y}+\cdots .&&\label{eq:one}
\end{flalign}}

Equation~\eqref{eq:one} is read term by term as Eq.~\eqref{eq:energy}. As the second derivative of the energy, $H_{\rm eff}$ is a Hessian, that is, a quadratic form on the tensor at site $i$: %
\definecolor{hred}{HTML}{D9534F}\definecolor{hrede}{HTML}{A33A37}%
\definecolor{hyel}{HTML}{F2C14E}\definecolor{hyele}{HTML}{B88A1E}%
\newcommand{\hballR}{\tikz[x=\hu,y=\hu,baseline=-0.6ex]\path[fill=hred,draw=hrede,line width=0.5pt] (0,0) circle (1.2);}%
\newcommand{\hballY}{\tikz[x=\hu,y=\hu,baseline=-0.6ex]\path[fill=hyel,draw=hyele,line width=0.5pt] (0,0) circle (1.2);}%
\newcommand{\hnetFfill}{\vcenter{\hbox{\begin{tikzpicture}[x=1.1\hu,y=1.1\hu]%
 \useasboundingbox (-5.2,-9.8) rectangle (35.3,9.8);%
 \foreach \y in {3.7,-3.7}{\draw[hline] (-2.3,\y)--(0,\y);\draw[hline] (30.1,\y)--(32.4,\y);\draw[hdot] (-5,\y)--(-3.2,\y);\draw[hdot] (33.3,\y)--(35.1,\y);%
   \draw[hline] (0,\y)--(2.6,\y);\draw[hline] (26.7,\y)--(30.1,\y);}%
 \hcolop{0}{0}{X}\hcolop{30.1}{0}{Y}%
 \hbrl{2.6}{0}{8.6}\hbrr{26.7}{0}{8.6}%
 \node[font=\scriptsize] at (28.2,-9.3) {$12$};%
 \node at (6.4,0) {\Large $F_J$};%
 \hparl{9.6}{0}{7.7}{1.9}\hparr{22.9}{0}{7.7}{1.9}%
 \foreach \y in {2,-2}{\draw[hline] (11.9,{3.9+\y})--(15.1,{3.9+\y}); \draw[hline] (17.1,{3.9+\y})--(20.3,{3.9+\y});%
   \draw[hline] (17.1,{-3.9+\y})--(20.3,{-3.9+\y});}%
 \hcolc{13.5}{3.9}%
 \draw[hline] (18.7,2.85)--(18.7,4.95);%
 \path[fill=hyel,draw=hyele,line width=0.5pt] (18.7,5.9) circle (0.95);%
 \path[fill=hred,draw=hrede,line width=0.5pt] (18.7,1.9) circle (0.95);%
 \hcolc{18.7}{-3.9}%
 \node at (13.5,-3.9) {\large $0$};%
\end{tikzpicture}}}}%
filling the open slot $T(\slot,\slot)$ (Appendix~\ref{sec:open-slot}) with a bra tensor (red) and a ket tensor (yellow) gives a number,
\begin{equation*}
 \bra{\hballR}H_{\rm eff}\ket{\hballY}=\bra{L_X}F_J^{[1]}[\Tconn]\bigl(T(\hballR,\hballY)\bigr)\ket{R_Y}+\cdots .
\end{equation*}
iDMRG and VUMPS need only the Hessian-vector map, which fills only the ket of the open slot,
\begin{equation*}
 H_{\rm eff}A_i=\bra{L_X}F_J^{[1]}[\Tconn]\bigl(T(\slot,A_i)\bigr)\ket{R_Y}+\cdots ;
\end{equation*}
their eigensolvers use only this action, so $H_{\rm eff}$ is never explicitly stored~\footnote{In a practical implementation, the fixed intermediate quantities in these expressions should be cached.}. The other terms are of the same kind: with the open slot on $X$ or $Y$ or outside the interaction string, they are matrix functions of the transfer matrix applied to a vector, as in the energy itself. Once the sum over the terms of the Hamiltonian is written as a function of a block matrix, every term of $H_{\rm eff}$ can be evaluated without a sum over positions and without forming any matrix (Appendix~\ref{sec:matrix-free}).

The derivation above uses the uniform tensor $A$ on every site. iDMRG and VUMPS instead work in the mixed canonical form, which differs from the uniform form only by a gauge transformation. They also need the actions on the bond matrix and on a two-site center, which follow from the same rule with a different open slot. Appendix~\ref{sec:mixed} gives these forms; everything else in iDMRG and VUMPS is unchanged.

\subsection{Energy variance}
\label{sec:variance-method}
The energy variance measures how far a state is from an eigenstate: it vanishes only for an exact eigenstate. It is commonly used to extrapolate the energy to the zero-variance limit in variational Monte Carlo~\cite{Sorella2001}, MPS~\cite{McCulloch2008,Michel2010,Hubig2018Variance}, and infinite projected entangled pair states (iPEPS)~\cite{CortesEstay2026}. Unlike finite-entanglement scaling~\cite{Tagliacozzo2008,Pollmann2009} and finite-correlation-length scaling~\cite{Corboz2018}, which are derived for critical states whose correlation length diverges with $\chi$, the variance extrapolation also applies to states that may be gapped.
The energy variance per physical site is
\begin{equation}
 v=\lim_{L\to\infty}\frac{\langle H_L^2\rangle-\langle H_L\rangle^2}{L}.
 \label{eq:variance}
\end{equation}
For $H=\sum_i\sum_{d\geq1}J_dX_iY_{i+d}+\cdots$, translation invariance gives
\begin{equation}
 \begin{split}
 v=\sum_{d_1,d_2\geq1}J_{d_1}J_{d_2}\sum_{j\in\mathbb Z}\Bigl[&\langle h_{0,d_1}h_{j,j+d_2}\rangle\\
 &-\langle h_{0,d_1}\rangle\langle h_{j,j+d_2}\rangle\Bigr]+\cdots,
 \end{split}
 \label{eq:variance-sum}
\end{equation}
with $h_{i,j}=X_iY_j$.
Once the order of the four sites $0,d_1,j,j+d_2$ along the chain is fixed, each term is a string of transfer matrices with operator insertions. The sums over $d_1$, $d_2$, and $j$ are carried out by contractions of the same kind as for the energy (Appendix~\ref{sec:matrix-free}). No cutoff is placed on $d_1$, $d_2$, or $j$.

The variance is computed as the difference of two large terms. As the state approaches an eigenstate, these contributions nearly cancel, causing a loss of numerical precision~\cite{Hubig2018Variance}. This limits the smallest energy variance per site that can be reliably resolved (Sec.~\ref{sec:coulomb}).
\section{The half-filled lattice Coulomb chain}
\label{sec:coulomb}
\label{sec:variance-groundstates}
\subsection{Model and method}
The lattice Coulomb chain is the simplest model in which the unscreened $1/d$ interaction competes with kinetic energy and lattice commensurability. We study this model for two reasons. First, we aim to provide high-precision reference energies for a model with long-range interactions. Second, we ask whether it has a measurable CDW order at weak coupling, where commensurability should pin a CDW whose amplitude is extremely small~\cite{Capponi2000,Valenzuela2003} and earlier numerics did not resolve it (Sec.~\ref{sec:introduction}). Resolving such a small order parameter requires systematic control of both interaction and variational errors. Here, we treat the $1/d$ interaction exactly, without an SOE fit, and perform extrapolation in the energy variance, both of which TMC provides. We consider spinless electrons on a one-dimensional lattice
\begin{equation}
 H_\rho=-\sum_i(c_i^\dagger c_{i+1}+\mathrm{h.c.})
 +V\sum_{i<j}\frac{(n_i-\rho)(n_j-\rho)}{(j-i)^\alpha},
 \label{eq:model}
\end{equation}
with $n_i=c_i^\dagger c_i$, the interaction strength $V$, the density $\rho=1/2$, and $\alpha=1$, the exponent of the unscreened lattice Coulomb tail. The hopping amplitude is the unit of energy. The density deviations $n_i-\rho$ incorporate a uniform neutralizing background.

As in Sec.~\ref{sec:calculus}, the expectation value of each interaction term, with the ends $X=Y=n-\rho$, splits into the disconnected part $\langle X\rangle\langle Y\rangle$ and a connected part. Summed over the distance $d$, the connected parts give the connected interaction energy per site $e_J^{\rm c}=\bra{L_X}F_J(\Tconn)\ket{R_Y}$ of Eq.~\eqref{eq:energy}. Without the neutralizing background, the ends would be $X=Y=n$, and the disconnected parts would contain $\rho^2\sum_dJ_d=\rho^2F_J(1)$, which diverges: $F_J(z)=-V\ln(1-z)/z$ is singular at $z=1$, the dominant eigenvalue of $T$. The projectors already keep the energy finite (Sec.~\ref{sec:energy-gradient}); charge conservation also removes the singularity from $F_J$ itself, so that rounding errors along $\ket{R}$ are not amplified: with $\Uone$-symmetric tensors and the neutralizing background, each end of a density deviation $n_i-\rho$ contributes a factor $\id-T$, which replaces $F_J(z)$ by $(1-z)^2F_J(z)$, whose coefficients are the second difference of the interaction in the distance (Appendix~\ref{sec:ward}).

The disconnected part contains no string of transfer matrices. It vanishes if $\langle n_i\rangle=\rho$ on every site. The CDW state with a two-site unit cell is an example in which it does not: $\langle n_i-\rho\rangle=(-1)^im_{\rm CDW}$, where $m_{\rm CDW}$ is the CDW amplitude, so the disconnected parts \mbox{$J_d\langle n_i-\rho\rangle\langle n_{i+d}-\rho\rangle$} of the terms that start at a given site add up to the alternating series
\begin{equation*}
 \sum_{d\geq1}(-1)^dJ_d\,m_{\rm CDW}^2=-F_J(-1)\,m_{\rm CDW}^2=-V\ln2\,m_{\rm CDW}^2 .
\end{equation*}
This contribution depends only on the local densities within the unit cell, so it is a local term that is added to the connected energy per site.

\subsection{Results}
We compute ground states of Eq.~\eqref{eq:model} at half filling for $V=4$, where the CDW order is unambiguous and little extrapolation is needed, so that the variance extrapolation can be tested, and for $V=2$, where the amplitude is very small. We use $\Uone$-symmetric iDMRG and VUMPS~\cite{Singh2010,Singh2011,McCulloch2008,Zauner2018} with the two sites of the CDW unit cell merged into one tensor of local physical dimension 4.
Both solvers use the same effective operators, with $F_J$ and its derivatives in the open-slot sums evaluated by Krylov matrix functions (Appendix~\ref{sec:matrix-free}). The Krylov dimension is increased until the estimated relative change in both the left and right connected environments is at most $10^{-9}$. We estimate this change by comparing the approximation of Krylov dimension $n_{\rm K}$ with one from a slightly smaller Krylov space $n_{\rm K}-24$; the criterion was met below a cap of $n_{\rm K}=1024$ throughout. %
We use bond dimensions up to $\chi=3072$ at $V=2$ and up to $\chi=1024$ at $V=4$. The complete optimization workflow that produced all $V=2$ states up to $\chi=3072$ took less than a day on one die of an AMD MI250X GPU on LUMI-G. %

We measure the energy per site $e$, the CDW amplitude $m_{\rm CDW}=|\langle n_A\rangle-\langle n_B\rangle|/2$ with $A$ and $B$ the two sites of the unit cell, and the energy variance per site $v$ of Eq.~\eqref{eq:variance}.
We extrapolate the energy and the CDW amplitude to $v=0$ (Sec.~\ref{sec:variance-method}). Theory predicts a gapped CDW state at every $V>0$~\cite{Capponi2000,Valenzuela2003}, so the scalings in $\chi$ derived for critical states~\cite{Tagliacozzo2008,Pollmann2009} do not apply here, whereas the variance extrapolation also applies to gapped states. For the energy we use the linear form $e=e_0+av$, which is typically expected for the energy as $v\to0$~\cite{Hubig2018Variance}, and for the CDW amplitude the offset power law $m_{\rm CDW}=m_0+bv^p$, which we compare with a pure power law below. Here $e_0$, $a$, $m_0$, $b$, and $p$ are fit parameters.

\begin{figure}[t]
\centering
\includegraphics[width=\columnwidth]{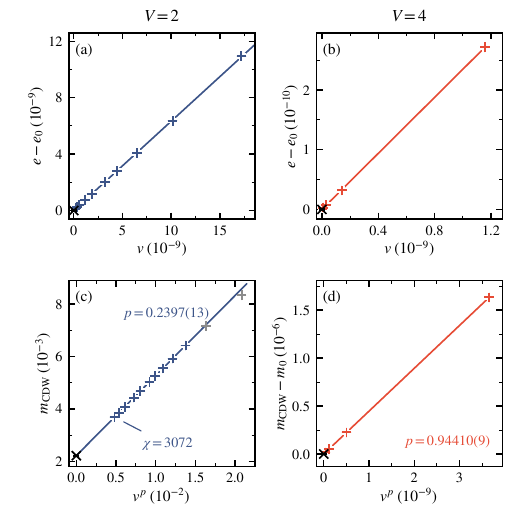}
\caption{The energy and CDW amplitude for the half-filled Coulomb chain, extrapolated to zero variance for $V=2$ (left) and $V=4$ (right).
(a,b) Energy per site $e$ as a function of the energy variance $v$, with the fit $e=e_0+av$ and the extrapolated values $e_0=-0.863815982983(15)$ and $-1.1121016662500(7)$, respectively. (c,d) CDW amplitude $m_{\rm CDW}$ as a function of $v^p$, with the fit $m_{\rm CDW}=m_0+bv^p$ and the extrapolated values $m_0=0.002213(14)$ and $0.09923211059(3)$, respectively. In panels (a), (b), and (d), the extrapolated values $e_0$ and $m_0$ are subtracted. The fitted exponent $p$ is given in panels (c) and (d), and the point with $\chi=3072$ is marked in panel (c). Plus signs show the numerical results, and black crosses with error bars indicate the extrapolated values. Gray symbols ($\chi=256$ and 384 at $V=2$) are not included in the fit and lie outside the range of panel (a).
Lines show fits to the data for $\chi=512$--3072 at $V=2$ and 256--640 at $V=4$.
Numbers in parentheses indicate $1\sigma$ fit uncertainties in the last quoted digits.}
\label{fig:highchi}
\end{figure}
\begin{figure}[t]
\centering
\includegraphics[width=\columnwidth]{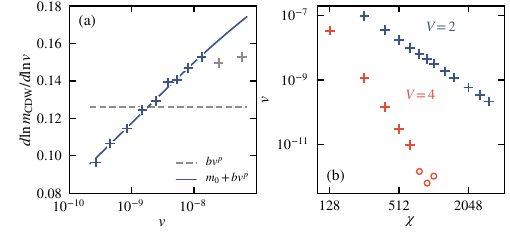}
\caption{Local slope of the CDW amplitude and energy variance of the Coulomb chain. (a) Local slope $d\ln m_{\rm CDW}/d\ln v$ at $V=2$. The slope is constant for a pure power law $m_{\rm CDW}=bv^p$, whereas a slope that decreases toward smaller $v$ is consistent with a nonzero $m_0$. Plus signs show the local slopes between neighboring states, in blue where both states lie in the fitting range and in gray where at least one lies below it. The dashed line and the blue line show the logarithmic slopes calculated from the fitted functions $m_{\rm CDW}=bv^p$ and $m_{\rm CDW}=m_0+bv^p$, respectively, both fitted to the data for $\chi=512$--3072.
(b) Energy variance as a function of $\chi$, in blue at $V=2$ and in coral at $V=4$. Open circles indicate variance estimates at the numerical cancellation floor; the corresponding states are excluded from the extrapolations.}
\label{fig:site}
\end{figure}
\begin{table}[t]
\centering\footnotesize
\renewcommand{\arraystretch}{1.18}
\begin{tabular*}{\columnwidth}{@{\extracolsep{\fill}}l|cc}
\toprule
 & $V=2$ & $V=4$\\
\midrule
fitting range of $\chi$ & 512--3072 & 256--640\\
\midrule
$e(\chi_{\max})$ & $-0.863815982822$ & $-1.112101666247$\\
$v(\chi_{\max})$ & $2.1\times10^{-10}$ & $9.3\times10^{-12}$\\
$m_{\rm CDW}(\chi_{\max})$ & $0.003685$ & $0.09923213$\\
\midrule
$e_0$ & $-0.863815982983(15)$ & $-1.1121016662500(7)$\\
$e(\chi_{\max})-e_0$ & $1.6\times10^{-10}$ & $3.1\times10^{-12}$\\
$m_0$ & $0.002213(14)$ & $0.09923211059(3)$\\
$p$ & $0.2397(13)$ & $0.94410(9)$\\
\bottomrule
\end{tabular*}
\caption{Data at the largest bond dimension and zero-variance extrapolation for the half-filled Coulomb chain. The first row gives the fitting range of $\chi$. The second block lists the energy per site $e$, the variance $v$, and the CDW amplitude $m_{\rm CDW}$ at the largest bond dimension $\chi_{\max}$ of this range. The third block lists the extrapolated values $e_0$ and $m_0$, the difference $e(\chi_{\max})-e_0$, and the exponent $p$. The intercept $e_0$ is from a fit of $e$ linear in $v$, and $m_0$ and $p$ are the intercept and exponent of the fit $m_{\rm CDW}=m_0+b\,v^p$. Numbers in parentheses indicate $1\sigma$ fit uncertainties in the last quoted digits. They are conditional on the fit form and the fitting window and do not include systematic errors.}
\label{tab:benchmark}
\end{table}
At $V=4$ the extrapolation gives the CDW amplitude $m_0=0.09923211059(3)$ (Fig.~\ref{fig:highchi}, Table~\ref{tab:benchmark}); the CDW amplitude obtained at $\chi=640$ already agrees with the extrapolated value to six significant digits [Fig.~\ref{fig:highchi}(d)]. The energy of this state lies $3\times10^{-12}$ above the extrapolated $e_0$. An extrapolation of exact diagonalization (ED) energies of finite chains (Appendix~\ref{sec:ed-alpha-one}) agrees with $e_0$ to within $2.1\times10^{-4}$. The variance falls by three and a half orders of magnitude, from $3\times10^{-8}$ at $\chi=128$ to $9.3\times10^{-12}$ at $\chi=640$ [Fig.~\ref{fig:site}(b)]. We exclude the states with $\chi\geq768$, whose variances are too small to be reliably resolved because of cancellation between large contributions.

At $V=2$ the energy is also well converged ($1.6\times10^{-10}$ above $e_0$ at $\chi=3072$) and agrees with the ED extrapolation to within $2.1\times10^{-4}$, as at $V=4$. The CDW amplitude is small and keeps decreasing, from $0.0083$ at $\chi=256$ to $0.0037$ at $\chi=3072$, while the correlation length of the iMPS grows to the very large value of about $2\times10^4$ sites and the variance is still decreasing.

To distinguish between a vanishing and a finite CDW amplitude as $v\to0$, we examine the local logarithmic slope $d\ln m_{\rm CDW}/d\ln v$ [Fig.~\ref{fig:site}(a)]. A pure power law $m_{\rm CDW}=bv^p$ has the constant slope $p$, whereas $m_{\rm CDW}=m_0+bv^p$ has the slope $pbv^p/(m_0+bv^p)$, which tends to zero as $v$ decreases when $m_0>0$. The measured slope decreases from $0.15$ to $0.10$.
A fit of $m_0+bv^p$ to the data for $\chi=512$--3072 reproduces this decrease of the slope and gives $p=0.2397(13)$ and $m_0=0.002213(14)$. States with bond dimensions below this range (gray) depart from this fit.
The data therefore favor $m_0>0$, a small nonzero CDW order. Such a small CDW amplitude is consistent with the weak lattice pinning anticipated by theory~\cite{Capponi2000,Valenzuela2003}, but was not resolved in earlier numerical studies, including DMRG calculations using an SOE with $K=6$ exponentials, open chains of up to $N=1600$ sites, and bond dimensions up to 500~\cite{Li2019}. %

\section{Excitations and the nonanalytic spectrum}
\label{sec:excitations}
This section applies TMC to excitations. Within the framework of the single-site excitation ansatz~\cite{Haegeman2012}, we use TMC to construct the excitation action and apply it to an exactly solvable model to compare the excitation spectrum of an algebraic interaction with that of its SOE fit.

\subsection{Excitation action}
\label{sec:exc-action}
Excitations above an iMPS ground state with one-site tensor $A$ (Sec.~\ref{sec:calculus}) are described variationally in its tangent space. The single-site excitation ansatz~\cite{Haegeman2012,Haegeman2013Tangent,Vanderstraeten2019} is $\ket{\Phi_k(B)}=\sum_j\e^{\ii kj}\ket{\Psi_j(B)}$, a momentum-$k$ superposition of the states $\ket{\Psi_j(B)}$ in which the tensor $B$ replaces $A$ at site $j$:
\begin{flalign*}
 &\ket{\Phi_k(B)}=\sum_j\e^{\ii kj}\,\hnetansatz .&&
\end{flalign*}
Blue tensors are the ground-state iMPS tensor $A$, the same on every site, and the purple tensor is $B$.
The excitation action is the effective Hamiltonian $H_{\rm eff}(k)$ acting on the tensor $B$ of this ansatz; its eigenvalue problem gives the excitation energies $\omega(k)$. As for $H_{\rm eff}$ in Sec.~\ref{sec:effective-operators}, this action is the Hamiltonian contraction differentiated with respect to the bra tensor at site $0$, which leaves an open slot (dashed circle), with the tangent tensor $B$ in the ket layer at site $d$; the result is summed over $d$ with the phase $\e^{\ii kd}$:
\begin{flalign*}
 &H_{\rm eff}(k)B=\sum_{d\in\mathbb Z}\e^{\ii kd}\,\hnetexcblob&&
\end{flalign*}
In the MPO representation, the shaded Hamiltonian $H$ is replaced by a chain of MPO tensors $W$:
\begin{flalign*}
 &{}=\sum_{d\in\mathbb Z}\e^{\ii kd}\,\hnetexcmpo&&
\end{flalign*}
Expanding the Hamiltonian into its terms instead, we consider a term whose ends $X$ and $Y$ enclose the open slot and $B$, with $a$, $b$, and $c$ intervening transfer matrices. The interaction endpoints are separated by $a+b+c+3$ lattice spacings, giving the weight $J_{a+b+c+3}$. The open slot and $B$ are separated by $d=b+1$ lattice spacings, giving the phase $\e^{\ii k(b+1)}$:
\begin{flalign*}
 &{}=\sum_{a,b,c\ge0}J_{a+b+c+3}\e^{\ii k(b+1)}&&\\*[-2pt]
 &\qquad\hnetexcsum+\cdots&&
\end{flalign*}
The remaining terms ($+\cdots$) contain the opposite ordering, the open slot and $B$ on the same site, and terms with the open slot or $B$ at or outside an end. The sum over $a$, $b$, and $c$ is one block of $F_J$ applied to a $3\times3$ block-triangular matrix:
\begin{flalign*}
 &{}=\hnetexcF+\cdots .&&
\end{flalign*}
With the tangent transfer matrices $T(\slot,A)$ and $T(A,B)$ and momentum phase $\e^{\ii k}$ per unit translation, this matrix is
\begin{equation*}
 \mathbb T_k=\begin{pmatrix}
  T&T(\slot,A)&0\\
 0&\e^{\ii k}T&\e^{\ii k}T(A,B)\\
 0&0&T
 \end{pmatrix},
\end{equation*}
and its upper-right block collects the interaction strings containing both the open slot and $B$, with $B$ to the right of the open slot:
\begin{equation}
 \begin{aligned}
  &[F_J(\mathbb T_k)]_{13}\\
  &=F_J^{[2]}[T,\e^{\ii k}T,T]\bigl(T(\slot,A),\e^{\ii k}T(A,B)\bigr)\\
  &=\sum_{a,b,c\ge0}J_{a+b+c+3}\,\e^{\ii k(b+1)}\,T^aT(\slot,A)T^bT(A,B)T^c .
 \end{aligned}
  \label{eq:two}
\end{equation}
Equation~\eqref{eq:two} follows from the second identity in Eq.~\eqref{eq:fn-notation}, with $f=F_J$ and diagonal blocks $T$, $\e^{\ii k}T$, and $T$. The remaining terms are obtained analogously, with the block structure adapted to the relative positions of the open slot, $B$, and the interaction endpoints.

We end this subsection with a few technical details. Before $F_J$ is evaluated, each end operator is split into its ground-state mean and a part with zero mean, $X=\langle X\rangle+\delta X$. The mean parts give one-site terms with weight $F_J(1)=\sum_{d\geq1}J_d$, finite for $\alpha>1$, and a constant that cancels in $H-E_0$, with $E_0$ the ground-state energy, so the rewrite is exact and, as in Eq.~\eqref{eq:energy}, only $\Tconn$ acts next to the ends.
The remaining terms, the energy subtraction, and the projection onto the left gauge of $B$, which removes the directions of $B$ that leave $\ket{\Phi_k(B)}$ unchanged~\cite{Haegeman2013Tangent,Vanderstraeten2019}, complete $H_{\rm eff}(k)$. The excitation energies $\omega(k)$ are obtained by solving the generalized eigenvalue problem $H_{\rm eff}(k)B=\omega(k)\,N(k)B$, where $N(k)$ is the Gram matrix of the states $\ket{\Phi_k(B)}$.

\subsection{Excitations of a long-range hopping model}
\label{sec:exc-hopping}
Previous long-range MPS quasiparticle calculations have represented the interaction by an SOE~\cite{Vanderstraeten2018}. An algebraic tail, however, has a Fourier transform that is nonanalytic at $k=0$, whereas that of any fixed SOE is analytic, so a fit rounds off the excitation spectrum at small momentum. Such a cusp at $k=0$ appears in the magnon dispersion of trapped-ion spin chains with power-law couplings~\cite{Kranzl2023}. TMC keeps $F_J$ and thus the algebraic tail. To separate this effect from the finite-$\chi$ error of the ground state, we take an excitation above a product state that is an exact ground state; the nonanalyticity then comes from the algebraic hopping alone, not from criticality.

We consider the Hamiltonian
\begin{equation}
 H_{\rm lr}=h\sum_i n_i-\sum_{i,d\geq1}\frac{c_i^\dagger c_{i+d}+\mathrm{h.c.}}{d^2},
 \label{eq:mag-model}
\end{equation}
which describes free fermions with inverse-square hopping; here $n_i=c_i^\dagger c_i$ and the chemical potential $h>\pi^2/3$. Every one-particle energy is at least $h-\pi^2/3>0$, so the empty state is the ground state. This ground state is an exact product state, an iMPS with $\chi=1$. %

Its one-particle excitation has the exact dispersion
\begin{equation}
 \omega(k)-\omega(0)
 =2\sum_{d\geq1}\frac{1-\cos(kd)}{d^2}
 =\pi|k|-\frac{k^2}{2},\qquad |k|\leq\pi ,
 \label{eq:cusp}
\end{equation}
a cusp with $[\omega(k)-\omega(0)]/|k|\to\pi$ when $|k|\to 0$. For absolutely summable hopping amplitudes $J_d$, the dispersion is $\omega(k)-\omega(0)=2[\hat J(0)-\hat J(k)]$, where $\hat J(k)=\sum_{d\geq1}J_d\cos(kd)$. For $J_d=1/d^2$, $\hat J(k)=\pi^2/6-\pi|k|/2+k^2/4$ for $|k|\leq\pi$~\cite{DLMFFourierSeries}, which gives Eq.~\eqref{eq:cusp}.
If instead $J_d=u^TM^{d-1}v$ with all eigenvalues of $M$ of modulus below one, which includes any sum of $K$ exponentials, $J_K(d)=\sum_{j=1}^Kc_j\lambda_j^d$ with $|\lambda_j|<1$, then
\begin{equation}
 \hat J(k)=\mathrm{Re}\bigl[u^T\e^{\ii k}(\id-\e^{\ii k}M)^{-1}v\bigr]
\end{equation}
is $\mathrm{Re}[\e^{\ii k}F_J(\e^{\ii k})]$ with a rational $F_J$ (Sec.~\ref{sec:calculus}) and is therefore analytic and even in $k$, so $\hat J(k)=\hat J(0)+O(k^2)$ and
\begin{align}
 \omega_K(k)-\omega_K(0)&=k^2\sum_{d\geq1}d^2J_K(d)+O(k^4),\notag\\
 \lim_{k\to0}\frac{\omega_K(k)-\omega_K(0)}{|k|}&=0 .
 \label{eq:soe-analytic}
\end{align}
A fixed SOE therefore rounds off the cusp at small $k$.

\begin{figure}[t]
\centering
\includegraphics[width=\linewidth]{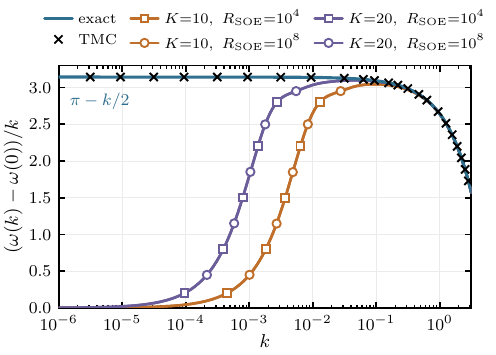}
\caption{Dispersion slope $[\omega(k)-\omega(0)]/k$ of the one-particle excitation above the exact product ground state of the Hamiltonian in Eq.~\eqref{eq:mag-model}. A cusp, $\omega(k)-\omega(0)\propto|k|$, gives a nonzero limit of this ratio for $k\to0^+$, whereas a dispersion that is analytic and even in $k$ gives the limit zero. The blue line is the exact value $[\omega(k)-\omega(0)]/k=\pi-k/2$ for $k>0$, and the crosses are the TMC values for $10^{-6}\leq k/\pi\leq0.9$. The four SOE curves show the dispersion of the same model with the hopping $d^{-2}$ replaced by an SOE with $K$ exponentials, fitted on all integer distances $1\leq d\leq R_{\mathrm{SOE}}$ with the TeNPy 1.1.0 routine~\cite{TenpyFit110}. Orange and purple curves have $K=10$ and $20$, and open squares and circles mark $R_{\mathrm{SOE}}=10^4$ and $10^8$, respectively.}
\label{fig:excitation}
\end{figure}

We confirm this analysis numerically. Figure~\ref{fig:excitation} shows that TMC reproduces the exact cusp, whereas every SOE fit of the same hopping rounds it off at small momentum. TMC evaluates the excitation action above with the exact generating function $F_J(z)=\Li_2(z)/z$ of the inverse-square hopping, and at all sampled momenta, $10^{-6}\leq k/\pi\leq0.9$, its value of $\omega(k)-\omega(0)$, with $\omega(0)$ from the exact solution, differs from Eq.~\eqref{eq:cusp} by less than $5\times10^{-15}$.
The SOE curves are the exact Fourier sums of the TeNPy fits~\cite{Hauschild2018Tenpy,TenpyFit110,Pirvu2010}, and they depart from the cusp below a momentum scale that is smaller for $K=20$ than for $K=10$; a larger maximum fitting distance $R_{\mathrm{SOE}}$ does not reduce this scale.
\section{First-quantized iMPS}
\label{sec:firstq}
This section shows that TMC also applies to a first-quantized iMPS, which uses one tensor per particle rather than per lattice site. We extend the first-quantized MPS of Li and Waintal~\cite{LiWaintal2026}, formulated for finite chains, to the thermodynamic limit. As a test case, we study the half-filled Coulomb chain of Eq.~\eqref{eq:model}, for which Sec.~\ref{sec:coulomb} provides second-quantized reference results.

In the first-quantized iMPS, the physical indices encode the distances between neighboring particles (the gaps). Hence, the distance between two particles is the sum of the gaps between them and is not determined by the number of tensors between them, so the generating-function construction of Sec.~\ref{sec:calculus}, in which the power of the transfer matrix counts the distance, does not apply directly. Nevertheless, the physical distance between particles can be reconstructed from their separation along the tensor chain and the $\Uone$ charges~\footnote{This $\Uone$ symmetry is not the one that follows from particle-number conservation in the standard sense: the conserved charge is the total length $\sum_iq_i$, the sum of the gaps between neighboring particles (Sec.~\ref{sec:firstq-u1}). We emphasize that it offers a different understanding of how a symmetry arises in a tensor network.} on the corresponding virtual bonds. This allows the interaction to be expressed using functions of the first-quantized iMPS transfer matrix (Sec.~\ref{sec:firstq-u1}). The algebraic interaction can therefore be evaluated without a cutoff in the distance or an SOE fit, directly in the thermodynamic limit.

Section~\ref{sec:firstq-gap} introduces the gap representation.
Section~\ref{sec:firstq-energy} writes the energy as a function of a matrix that advances the distance by one site. Section~\ref{sec:firstq-u1} reduces this matrix to the transfer matrix of the particle chain for an integer mean gap and derives the effective Hamiltonians of iDMRG and VUMPS. Section~\ref{sec:firstq-compare} compares first-quantized with second-quantized states of the Coulomb chain at $V=2$ and $V=4$, and Sec.~\ref{sec:firstq-free} studies the finite-entanglement scaling at $V=0$.

\subsection{Gap representation}
\label{sec:firstq-gap}
In first quantization, a basis configuration is labeled by the ordered particle coordinates $x_1<x_2<\cdots$. Following the gap representation of Li and Waintal~\cite{LiWaintal2026},
we replace the coordinates by the positive gaps $q_i=x_i-x_{i-1}$, which fix the configuration up to an overall translation. In this basis each particle is assigned one integer $q_i\geq1$.
A quantum state in this basis is represented by an iMPS on the particle chain:
\begin{flalign*}
 &\ket{\Psi(A)}=\sum_{x_1<x_2<\cdots}\psi(x_1,x_2,\ldots)\,\ket{x_1,x_2,\ldots}&&\\
 &\phantom{\ket{\Psi(A)}}=\sum_{\{q_i\}}\,\hnetfqstate\,\ket{\ldots,q_{i-1},q_i,q_{i+1},\ldots}.&&
\end{flalign*}
Each blue tensor represents the same tensor $A$, with physical index $q_i$. The dots indicate that the chain extends infinitely in both directions.
In the variational calculation the physical index is restricted to $1\leq q_i\leq\gapmax$, where $\gapmax$ is the maximum gap between two neighboring particles allowed in this ansatz. 

\subsection{Energy in the gap basis}
\label{sec:firstq-energy}
Let $T_q=A^q\otimes\overline{A^q}$ and let $T=\sum_{q=1}^{\gapmax}T_q$ be the transfer matrix of the particle chain, as in Eq.~\eqref{eq:transfer-function} with the gap as physical index. We denote its fixed points by $\bra{L}$ and $\ket{R}$, with $\braket{L}{R}=1$. The hopping is local in the gap basis, because it changes only two neighboring gaps. The Coulomb interaction is not local in this basis, because the physical distance between two particles is the sum of all gaps between them. We therefore evaluate its energy using a recursion over the physical distance. We first give the gap distribution, which fixes the density, and then the kinetic and Coulomb energies per particle.

\emph{Gap distribution.} The probability of a gap $q$ is $P(q)=\bra{L}T_q\ket{R}$, and the mean gap $\mu=\sum_qqP(q)=\rho^{-1}$ fixes the density.

\emph{Hopping.} As discussed in Ref.~\cite{LiWaintal2026}, a hop of a particle by one lattice site changes only the two gaps next to it, with amplitude $-1$. In the diagram below, the top row shows the gaps $(q,p)$ before the hop and the bottom row the gaps $(q+1,p-1)$ after it:
\begin{equation*}
 \hnetfqhopii
\end{equation*}
In the gap basis the kinetic term is therefore a two-site operator $h$ on the gaps of neighboring tensors of the particle chain~\footnote{With $n_i=q_i-1\geq0$ empty sites between neighboring particles and without the cutoff, the first-quantized hopping of the free fermions is exactly the second-quantized phase model $-\sum_i(\phi_i^\dagger\phi_{i+1}+\mathrm{h.c.})$ on the chain of particles, with $\phi\ket{n}=\ket{n-1}$ and $\phi\ket{0}=0$, which is solvable by the Bethe ansatz~\cite{BogoliubovIzerginKitanine1998,Bogoliubov2005}. Free fermions in first quantization and the phase model in second quantization thus have the same Hamiltonian tensors, so that the entanglement entropy scaling of the first-quantized iMPS can also be regarded as a property of the phase model. This points to a correspondence between first- and second-quantized iMPS that remains to be studied.}, with matrix elements
\begin{equation*}
 h_{q^\prime p^\prime,qp}=-\bigl(\delta_{q^\prime,q+1}\delta_{p^\prime,p-1}+\delta_{q^\prime,q-1}\delta_{p^\prime,p+1}\bigr)
\end{equation*}
for $1\leq q^\prime,p^\prime\leq\gapmax$, and zero otherwise.
The kinetic energy per particle, $e_{\rm kin}$, is the expectation value of $h$:
\begin{flalign}
 &e_{\rm kin}=\hnetfqkin&&\notag\\
 &\phantom{e_{\rm kin}}=\sum_{q,p,q^\prime,p^\prime}h_{q^\prime p^\prime,qp}\bra{L}\bigl(A^q\otimes\overline{A^{q^\prime}}\bigr)\bigl(A^p\otimes\overline{A^{p^\prime}}\bigr)\ket{R}.&&
 \label{eq:firstq-kin}
\end{flalign}

\emph{Coulomb interaction.} For the interaction term of Eq.~\eqref{eq:model}, the Coulomb energy per particle is
\begin{equation}
 e_V=V\sum_{d\geq1}\frac{u_A(d)-\rho}{d^\alpha},
 \label{eq:firstq-eV}
\end{equation}
where $u_A(d)$ is the probability that a particle sits at distance $d$ from a given particle in the first-quantized iMPS with tensor $A$, and the density $\rho$ supplies the neutralizing background. The Coulomb energy per site is the particle density times the Coulomb energy per particle, $\rho e_V$.

The probability $u_A(d)$ is a sum over the arrangements of the particles in between: each gap sequence with $q_1+\cdots+q_k=d$ contributes $\bra{L}T_{q_1}\cdots T_{q_k}\ket{R}$. Both the number $k$ of tensors and the gaps vary, so the number of tensors does not fix the distance, and the sum is not a power of $T$. Grouping the sequences by their first gap shows that the vectors $\ket{R_d}=\sum_{q_1+\cdots+q_k=d}T_{q_1}\cdots T_{q_k}\ket{R}$ obey $\ket{R_d}=\sum_{q=1}^{\gapmax}T_q\ket{R_{d-q}}$, with $\ket{R_0}=\ket{R}$ and $\ket{R_d}=0$ for $d<0$, and $u_A(d)=\braket{L}{R_d}$. The stacked vector $(\ket{R_d},\ket{R_{d-1}},\ldots,\ket{R_{d-\gapmax+1}})^T$ is advanced from $d$ to $d+1$ by the matrix
\begin{equation}
 C=\begin{pmatrix}
 T_1&T_2&\cdots&T_{\gapmax}\\
 \id&0&\cdots&0\\
 &\ddots&\ddots&\vdots\\
 0&&\id&0
 \end{pmatrix},\qquad
 u_A(d)=\bra{\ell}C^d\ket{r},
 \label{eq:firstq-companion}
\end{equation}
with $\bra{\ell}=(\bra{L},0,\ldots,0)$ and $\ket{r}=(\ket{R},0,\ldots,0)^T$ the stacked vector at $d=0$; each factor of $C$ advances the distance by one site. The sum over arrangements is thus a power of $C$.

Subtracting the background $\rho$ projects the eigenvalue 1 of $C$, assumed nondegenerate, out of the distance sum, as for $T$ in Sec.~\ref{sec:calculus}: its contribution to every $u_A(d)$ is exactly $\rho$. With $\widetilde C$ the matrix $C$ with this eigenvalue projected out, Eq.~\eqref{eq:firstq-eV} becomes
\begin{equation}
 e_V=V\bra{\ell}\operatorname{Li}_\alpha(\widetilde C)\ket{r},\qquad \operatorname{Li}_\alpha(z)=\sum_{d\geq1}\frac{z^d}{d^\alpha},
 \label{eq:firstq-energy}
\end{equation}
which has the form of Eq.~\eqref{eq:energy} with $C$ in place of $T$. Equation~\eqref{eq:firstq-energy} is exact, but $C$ has dimension $\gapmax\chi^2$, since it stacks the $\gapmax$ most recent vectors $\ket{R_d}$. Its lower blocks only shift these vectors, so $C$ need not be stored and can be applied at the cost of $T$, but the Krylov vectors still have dimension $\gapmax\chi^2$. In the following subsection we explain how to avoid this additional cost by exploiting $\Uone$ symmetry.

\subsection{Evaluation with spatial $\Uone$ charge}
\label{sec:firstq-u1}
We describe the origin of the $\Uone$ symmetry of the first-quantized iMPS used in this paper and show how the virtual charges encode the distance between two particles. With $\Uone$ symmetry, the energy of Eq.~\eqref{eq:firstq-energy} can be computed directly from the transfer matrix $T$, of dimension at most $\chi^2$, without using the enlarged matrix~$C$.

\emph{Symmetric tensors.} A global $\Uone$ symmetry, generated by a conserved integer charge, can be built exactly into the tensors of an MPS~\cite{McCulloch2007,Singh2010,Singh2011}. Each basis state $a$ of the virtual bond has a charge $g_a$, and a tensor whose physical state $\sigma$ has the charge $c_\sigma$ conserves the charge element by element: $A^\sigma_{ab}\neq0$ only if $g_b-g_a=c_\sigma$. With $\hat Q$ the diagonal operator of the virtual charges, as in Appendix~\ref{sec:ward}, this reads
\begin{equation}
 [\hat Q,A^\sigma]=-c_\sigma A^\sigma.
 \label{eq:u1-tensor}
\end{equation}
 The tensors are then block sparse, and every contraction splits into charge sectors. In a uniform iMPS the physical charge is measured from its mean value per unit cell, so that the bond charges stay bounded along the infinite chain.

\emph{Second quantization.} The conserved charge is the particle number $N=\sum_xn_x$, and the physical state of a site is its occupation $n$. With $c_n=n-\rho$, Eq.~\eqref{eq:u1-tensor} reads
\begin{equation}
 [\hat Q,A^n]=-(n-\rho)A^n,
 \label{eq:u1-secondq}
\end{equation}
which is the Ward identity, Eq.~\eqref{eq:ward-commutator}, of Appendix~\ref{sec:ward}. Across the tensor of site $x$ the charge changes by $n_x-\rho$, so in each term of a contraction the bond to the right of site $x$ has the charge
\begin{equation*}
 g_x=g_0+\bigl[N(x)-\rho x\bigr],\qquad N(x)=\sum_{y=1}^xn_y .
\end{equation*}
Here $g_0$ is the charge on the bond to the right of site~$0$. Up to this constant, the bond charge records how many particles the left part holds in excess of the mean. The charge $g_x$ is an integer only when $\rho x$ is; at half filling the two sites of the CDW unit cell form one tensor (Sec.~\ref{sec:coulomb}), across which the charge changes by an integer.

\emph{First quantization.} Here the particle number is fixed by construction, one tensor per particle, so the particle number cannot serve as the $\Uone$ charge. Instead, the Hamiltonian conserves the total length $\sum_iq_i$: a hop changes two neighboring gaps by $+1$ and $-1$, and the Coulomb term is diagonal in the gaps.  The physical state of a tensor is its gap $q$, with charge $c_q=q-\mu$, where $\mu$ is the mean gap. Equation~\eqref{eq:u1-tensor} then reads
\begin{equation}
 [\hat Q,A^q]=-(q-\mu)A^q,
 \label{eq:gap-selection}
\end{equation}
which is Eq.~\eqref{eq:u1-secondq} with particles and sites exchanged: the tensor belongs to a particle, its physical index counts sites, and the mean gap $\mu$ replaces the density $\rho$~\footnote{The second-quantized charge is thus an Eulerian quantity, the excess particle number to the left of a fixed site, and the first-quantized charge a Lagrangian one, the displacement of a labeled particle [Eq.~\protect\eqref{eq:firstq-distance}]. The two are linked by $N(x_i)=i$, as in Haldane's harmonic-fluid description~\cite{Haldane1981}.}.

Written as a sum over virtual indices, a contraction has in each term, on every bond, a bond state of definite charge. Let $g_i$ be this charge on the bond at particle $i$, between the tensors of the gaps $q_i=x_i-x_{i-1}$ and $q_{i+1}$. By Eq.~\eqref{eq:gap-selection} a term is nonzero only if $g_i=g_{i-1}+(q_i-\mu)$ for every $i$. Summed along the chain, this gives $g_i=\delta_i$ up to a constant, where $\delta_i=x_i-\mu i$ is the displacement of particle $i$ from site $\mu i$ of a lattice with spacing $\mu$:
\begin{equation*}
 \hnetfqdisp
\end{equation*}
Particles $i$ and $i+s$ are therefore at distance
\begin{equation}
 d=x_{i+s}-x_i=\mu s+\delta_{i+s}-\delta_i=\mu s+g_{i+s}-g_i.
 \label{eq:firstq-distance}
\end{equation}
In this way, the $\Uone$ symmetry generated by the total length encodes the distance between two particles in the bond charges: the separation $s$ in particle labels gives the mean distance $\mu s$ (particles $i$ and $i+s$ have $s-1$ particles in between), and the charges on the two end bonds give the deviation from it, $g_{i+s}-g_i$.

Each term of $\bra{L}T_{q_1}\cdots T_{q_s}\ket{R}$ in $u_A(d)$ therefore has end charges $g$ and $g'$ with $q_1+\cdots+q_s=\mu s+g'-g$, and the sum over all gap sequences of total length $d$ becomes
\begin{equation}
 u_A(d)=\sum_{s\geq1}\ \sum_{g'-g=d-\mu s}\bra{L}\Pi_g\,T^s\,\Pi_{g'}\ket{R},
 \label{eq:firstq-uA-charge}
\end{equation}
where $\Pi_g$ projects the bond onto its bond states of charge $g$. The tensors between the two particles now enter only through $T^s$, without the stacked blocks of $C$ that hold the vectors $\ket{R_d}$ of the last $\gapmax$ distances; this replaces Eq.~\eqref{eq:firstq-companion}. Since $T$ shifts the ket and bra charges together, it acts on operators that are block diagonal in the charge, the charge-neutral sector, a space of dimension $\sum_g\chi_g^2\leq\chi^2$, with $\chi_g$ the number of bond states of charge $g$, instead of the $\gapmax\chi^2$ of $C$.

\emph{TMC evaluation.} We now write the distance sum as matrix functions of $T$ instead of $C$. In Eq.~\eqref{eq:energy} the number of transfer matrices between the two insertions fixes the distance, so the sum over distances weighted by the interaction $J_d$ is the matrix function $F_J(T)$. In first quantization the number of tensors does not fix the distance, but by Eq.~\eqref{eq:firstq-uA-charge} the end charges supply the remainder. Exchanging the order of the sums over the distance and over the number of tensors then turns the distance sum into matrix functions of $T$. We carry this out in three steps. 

First, inserting Eq.~\eqref{eq:firstq-uA-charge} gives a sum over $d$, $s$, and the end charges, constrained by $g'-g=d-\mu s$:
\begin{align}
 \sum_{d\geq1}\frac{u_A(d)}{d^\alpha}
 &=\sum_{d\geq1}\ \sum_{s\geq1}\ \sum_{g'-g=d-\mu s}\frac{\bra{L}\Pi_g\,T^s\,\Pi_{g'}\ket{R}}{d^\alpha}\nonumber\\
\intertext{Second, we exchange the order of summation. For given $s$, $g$, and $g'$, exactly one distance meets the constraint, $d=\mu s+g'-g$, so the sum over $d$ collapses. The distance then no longer labels the terms; it survives only in the weight, as a function of $s$ and of the charge difference $\Delta=g'-g$:}
 &=\sum_{g,g'}\ \sum_{s\geq1}\frac{\bra{L}\Pi_g\,T^s\,\Pi_{g'}\ket{R}}{(\mu s+g'-g)^\alpha}\nonumber\\
\intertext{Third, for fixed $g$ and $g'$ the sum over $s$ is a power series in $T$ with scalar coefficients, and thus defines a matrix function:}
 &=\sum_{g,g'}\bra{L}\Pi_g\,F_{\alpha,g'-g}(T)\,\Pi_{g'}\ket{R},
 \label{eq:firstq-eV-charge}
\end{align}
with
\begin{equation}
 \begin{split}
 F_{\alpha,\Delta}(z)&=\sum_{\substack{s\geq1\\\mu s+\Delta>0}}\frac{z^s}{(\mu s+\Delta)^\alpha}\\
 &=\mu^{-\alpha}z^{s_0}\,\Phi\bigl(z,\alpha,s_0+\Delta/\mu\bigr),
 \end{split}
 \label{eq:firstq-lerch-single}
\end{equation}
where $s_0$ is the smallest integer $s\geq1$ with $\mu s+\Delta>0$, and $\Phi(z,\alpha,\eta)=\sum_{j\geq0}z^j/(j+\eta)^\alpha$ is the Lerch transcendent. Terms with $\mu s+\Delta\leq0$ are left out because their matrix elements vanish: a nonzero term has $d=q_1+\cdots+q_s\geq s$.

Nothing in these steps depends on the power law. For an arbitrary pair interaction $J_d$ the same three steps give
\begin{equation}
 \begin{split}
 \sum_{d\geq1}J_d\,u_A(d)&=\sum_{g,g'}\bra{L}\Pi_g\,F_{J,g'-g}(T)\,\Pi_{g'}\ket{R},\\
 F_{J,\Delta}(z)&=\sum_{\substack{s\geq1\\\mu s+\Delta>0}}J_{\mu s+\Delta}\,z^s,
 \end{split}
 \label{eq:firstq-FJ}
\end{equation}
so that the interaction enters only through the scalar series $F_{J,\Delta}$, without a cutoff in the distance and without a fit of $J_d$. The exchange of the sums is term by term and holds whenever the series converges absolutely; for $\alpha=1$, where it does not, the background cancels the divergence, as shown below.

\emph{Implementation.} The function $F_{\alpha,\Delta}(z)$ is not analytic at $z=1$, the dominant eigenvalue of $T$, so we split $T=\ket{R}\bra{L}+\Tconn$ as in Sec.~\ref{sec:calculus}. Since $F_{\alpha,\Delta}(0)=0$, each term of Eq.~\eqref{eq:firstq-eV-charge} becomes
\begin{equation}
 \begin{split}
 \bra{L}\Pi_g\,F_{\alpha,g'-g}(T)\,\Pi_{g'}\ket{R}{}&=w_gw_{g'}F_{\alpha,g'-g}(1)\\
 &+\bra{L}\Pi_g\,F_{\alpha,g'-g}(\Tconn)\,\Pi_{g'}\ket{R},
 \end{split}
 \label{eq:firstq-split}
\end{equation}
where $w_g=\bra{L}\Pi_g\ket{R}$ is the probability that the bond carries charge $g$.
The first, disconnected term needs no matrix function: by Eq.~\eqref{eq:firstq-lerch-single} at $z=1$, $F_{\alpha,\Delta}(1)=\mu^{-\alpha}\zeta(\alpha,s_0+\Delta/\mu)$, a Hurwitz zeta function.
For $\alpha=1$ it diverges by the same amount for every $\Delta$. Unlike in Sec.~\ref{sec:coulomb}, where the background sits in the ends and removes the divergent part $\rho^2F_J(1)$, here the background is the term $-\rho$ of Eq.~\eqref{eq:firstq-eV}; since $\sum_gw_g=1$ and $\mu^{-1}=\rho$, it cancels the divergence, and a digamma function remains (Appendix~\ref{sec:ward}).
The second term is evaluated by a Krylov subspace method in the charge-neutral sector (Appendix~\ref{sec:matrix-free}).
Differentiating through $T$, the ends, and the eigenvectors gives the effective Hamiltonians of iDMRG and VUMPS, as in Sec.~\ref{sec:effective-operators}.

\subsection{Comparison with second-quantized states}
\label{sec:firstq-compare}
To compare with the second-quantized results of Sec.~\ref{sec:coulomb}, we evaluate the first-quantized energy for the model of Eq.~\eqref{eq:model} at half filling ($\mu=2$) with $\Uone$ symmetry and optimize the states with VUMPS.
We convert the first-quantized energy per particle to the energy per physical site via $E/L=\rho E/N$.
The reference values $e_0$ and $m_0$ are the zero-variance extrapolations of the second-quantized series of Sec.~\ref{sec:coulomb}.

\begin{figure}[t]
\centering
\includegraphics[width=\linewidth]{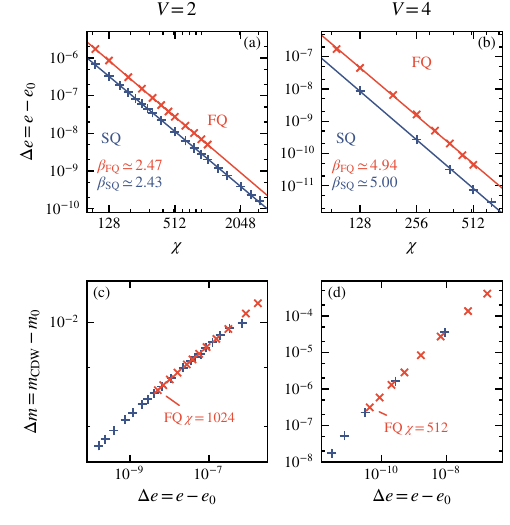}
\caption{Comparison of first-quantized (FQ) and second-quantized (SQ) states of the half-filled Coulomb chain ($\alpha=1$) at $V=2$ (left) and $V=4$ (right).
(a,b) Energy error per physical site $\Delta e=e-e_0$ as a function of the bond dimension $\chi$; lines show power-law fits to all displayed states of each sequence, with the powers $\beta$ given in the panels. (c,d) CDW error $\Delta m=m_{\rm CDW}-m_0$ as a function of the energy error.
Blue plus signs are SQ states and coral crosses are FQ states.
The FQ sequences reach $\chi=1024$ at $V=2$ and $\chi=512$ at $V=4$.
Reference values $e_0,m_0$ are the SQ zero-variance fits of Table~\ref{tab:benchmark}.}
\label{fig:firstq}
\end{figure}

Figure~\ref{fig:firstq} shows the results. At $V=2$ and $\gapmax=12$, the $\chi=1024$ first-quantized state has $e=-0.863815978$ and $m_{\rm CDW}=0.00571$, with energy error $5.0\times10^{-9}$.
At $V=4$ the $\chi=512$ state has $e=-1.112101666206$ and $m_{\rm CDW}=0.099232$, with energy error $4.4\times10^{-11}$. The gap cutoff $\gapmax=12$ is large enough not to affect these results~\footnote{Truncating the states to $\gapmax=10$ and restoring their canonical form by a gauge transformation changes their energy by less than $2\times10^{-13}$ per site.}.

For this model, to reach the same energy error, the first-quantized state needs a bond dimension larger by a nearly constant factor, about $1.45$ at $V=2$ and $1.42$ at $V=4$, as the nearly equal powers of the fits in Fig.~\ref{fig:firstq}(a,b) show ($\beta=2.47$ and $2.43$ at $V=2$, $4.94$ and $5.00$ at $V=4$, for FQ and SQ, respectively).
The two representations nevertheless approach the ground state along the same path: at equal energy error their CDW amplitudes agree to within $1.5\%$ [Fig.~\ref{fig:firstq}(c,d)].

\subsection{Finite-entanglement scaling at $V=0$}
\label{sec:firstq-free}
To compare the entanglement of the two representations at a critical point with a known central charge, we now turn to the same chain at $V=0$, i.e., free spinless fermions at half filling, which are critical with central charge $c=1$.
Finite-entanglement scaling of MPS predicts $\xi\propto\chi^\kappa$~\cite{Tagliacozzo2008,Pollmann2009}, and the half-chain entropy obeys
$S=a+(c/6)\ln\xi$~\cite{CalabreseCardy2004,Stojevic2015}. These laws are established for second-quantized MPS but not for the first-quantized iMPS of Ref.~\cite{LiWaintal2026}~\footnote{Nor has the meaning of the entropy at a bond of the first-quantized iMPS. A related but different entanglement between groups of particles is defined in Ref.~\cite{Zozulya2008}.}.

In both representations we compute the correlation length $\xi=-1/\ln|\lambda_2|$ from the second-largest eigenvalue $\lambda_2$ of the transfer matrix in the charge-neutral sector (Sec.~\ref{sec:firstq-u1}), which governs the correlations of charge-conserving operators such as the density, and the entanglement entropy $S=-\sum_j s_j^2\ln s_j^2$ from the Schmidt values $s_j$ at a bond~\cite{Vanderstraeten2019}. For the first-quantized iMPS, whose tensors are particles, $\xi$ is thus measured in transfer-matrix steps along the particle chain. The mean separation between neighboring particles sets the conversion to lattice units.

\begin{figure}[t]
\centering
\includegraphics[width=\columnwidth]{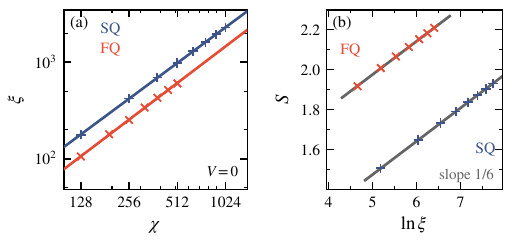}
\caption{Finite-entanglement scaling of free fermions at half filling, for second-quantized (SQ) states (blue plus signs) and first-quantized (FQ) states (coral crosses).
(a) Correlation length $\xi$ of the charge-neutral sector, in units of transfer-matrix steps (each step spans two lattice sites in SQ and advances the particle index by one in FQ), as a function of the bond dimension $\chi$. Lines show fits $\xi\propto\chi^\kappa$ to all displayed points, with $\kappa=1.228(3)$ for SQ and $1.261(8)$ for FQ.
(b) Entanglement entropy $S$ as a function of $\ln\xi$. Fits $S=a+(c/6)\ln\xi$ to all displayed points give $c=0.9951(6)$ for SQ and $1.001(5)$ for FQ. The gray lines have slope $1/6$ ($c=1$) and fitted offsets. Numbers in parentheses indicate $1\sigma$ fit uncertainties in the last quoted digits.}
\label{fig:firstq-free}
\end{figure}

Figure~\ref{fig:firstq-free} shows the results. Fits to all displayed points give $c=0.9951(6)$ for second-quantized states ($\chi=128$--1024) and $c=1.001(5)$ for first-quantized states ($\chi=128$--512, $\gapmax=12$), both close to $c=1$. The correlation-length powers are
$\kappa=1.228(3)$ and $1.261(8)$, both below the asymptotic value $\kappa\approx1.34$ for $c=1$~\cite{Pollmann2009}. The first-quantized data thus follow the same scaling forms, with an entropy slope consistent with $c=1$. Only the offsets differ, $a=0.65$ for SQ and $1.14$ for FQ.

\section{Discussion}
\label{sec:discussion}
We have established an MPO-free framework for variational iMPS simulations of chains with long-range interactions in the thermodynamic limit. A pair interaction enters only through its generating function $F_J$ acting on the iMPS transfer matrix, and the ground-state optimization, the tangent-space excitations, and the energy variance all follow from the same contraction rules. For interactions with an exact MPO, the framework gives the same results as the MPO approach; for interactions without an exact MPO, including algebraically decaying interactions, it treats the interaction directly, without an SOE fit. TMC can thus provide reference ground states and energy variances~\cite{Kokail2019} for benchmarking quantum simulations of one-dimensional systems with power-law couplings~\cite{Richerme2014,Jurcevic2014,Bernien2017,Saffman2010,Yan2013}.

For the half-filled Coulomb chain at $V=2$, we obtained highly accurate ground states with an energy variance per site as low as $2\times10^{-10}$, and extrapolation to zero variance indicated a small nonzero CDW amplitude, $m_0\approx0.0022$, consistent with the order predicted by theory. For an exactly solvable long-range hopping model, TMC reproduced the $|k|$ cusp of the one-particle dispersion, which any fixed SOE rounds off. For the Coulomb chain, we found that first- and second-quantized iMPS give nearly identical CDW amplitudes at equal energy error, with first-quantized iMPS requiring a bond dimension larger by a nearly constant factor over the range studied. At $V=0$, the entanglement entropy scaling in both representations is consistent with the expected central charge $c=1$.

Li and Waintal~\cite{LiWaintal2026} suggested that first-quantized MPS may be particularly efficient for the low-density Wigner-crystal regime. TMC makes it possible to test this by comparing first- and second-quantized iMPS directly in the thermodynamic limit. Extending TMC to finite MPS, which lack translation invariance, would require $F_J$ of a transfer matrix that varies along the chain. The extension of TMC to two dimensions with iPEPS is more challenging. The distance between two sites is no longer counted by powers of a single transfer matrix. Whether an analogous formulation can be developed using approximate iPEPS environments such as corner transfer matrices~\cite{NishinoOkunishi1996,OrusVidal2009} or boundary MPS~\cite{Jordan2008,Fishman2018} remains an open question.

More broadly, transfer matrices have been central to one-dimensional systems since Baxter's variational and corner transfer matrices~\cite{Baxter1968,Baxter1978}, through the transfer-matrix renormalization group to MPS~\cite{Nishino1995,NishinoOkunishi1996,Fannes1992,Ostlund1995,Verstraete2008,Cirac2021}; TMC turns this transfer-matrix view into an algorithmic framework for variational computations with long-range interactions.
 
\begin{acknowledgments}
We thank Hong-Hao Tu, Bei Qiao, Jheng-Wei Li, Zhen Yan, Yuchi He, and \mbox{J.-S.~Caux} for helpful discussions.
This project has received funding from the European Research Council under
the European Union's Horizon 2020 research and innovation program
(Grant Agreement No.\ 101001604).
We acknowledge the EuroHPC Joint Undertaking for awarding access to LUMI-G (Development Access, EHPC-DEV-2026D08-198) and Leonardo DCGP
(Benchmark Access, EHPC-BEN-2026B08-001). This work was also carried out on the Dutch national e-infrastructure with the support of SURF Cooperative.

\emph{AI disclosure.} Large language models (GPT-5-sol, GPT-6-sol, GPT-Astra, Claude Sonnet, Claude Opus, Claude Fable, Gemini-Flash, DeepSeek-V4-flash, and DeepSeek-V4.1-flash) assisted with coding, code execution, testing, and editing of the manuscript text. Q.~Y. reviewed the output and tested the code. Both authors are responsible for the final content.

\emph{Data availability.} The data and plotting scripts underlying the figures and tables are provided with the manuscript source package, and the TMC code will be made publicly available upon publication.
Our TMC implementation contains hardware-level performance optimizations developed with AI assistance, in particular GPU implementations of tensors with symmetries such as $\Uone$.
We acknowledge earlier implementations of symmetric tensors in ITensor, TensorKit.jl, Block2, and TeNPy~\cite{Fishman2022ITensor,TensorKit,Zhai2023Block2,Hauschild2018Tenpy}.
\end{acknowledgments}

\appendix

\section{Open slots}
\label{sec:open-slot}
\newcommand{\osCol}[1]{\vcenter{\hbox{\begin{tikzpicture}[x=3.7pt,y=3.7pt]%
 \definecolor{hbl}{HTML}{5271C4}\definecolor{hble}{HTML}{3A55A0}%
 \useasboundingbox (\ifnum#1=0 -5.2\else -6.8\fi,-7.8) rectangle (\ifnum#1=0 5.2\else 6.8\fi,7.8);%
 \draw[line width=1.3pt] (-2.6,3.7)--(2.6,3.7);%
 \ifnum#1=0
  \draw[line width=1.3pt] (-2.6,-3.7)--(2.6,-3.7);\draw[line width=1.3pt] (0,-2.5)--(0,2.5);%
  \path[fill=hbl,draw=hble,line width=0.5pt] (0,-3.7) circle (1.2);%
  \node[font=\scriptsize,inner sep=1pt] at (0,-6.4) {$\overline{A}$};%
 \else
  \draw[line width=1.3pt] (-4.2,-3.7)--(-1.5,-3.7);\draw[line width=1.3pt] (1.5,-3.7)--(4.2,-3.7);%
  \draw[line width=1.3pt] (0,-2.2)--(0,2.5);%
  \path[fill=white,draw=hble,line width=0.9pt,dash pattern=on 1.8pt off 1.3pt] (0,-3.7) circle (1.5);%
 \fi
 \ifnum#1=1
  \node[font=\scriptsize,inner sep=0.5pt,anchor=north east] at (-1.3,-5.0) {$\gamma'$};%
  \node[font=\scriptsize,inner sep=0.5pt,anchor=north west] at (1.3,-5.0) {$\delta'$};%
 \fi
 \path[fill=hbl,draw=hble,line width=0.5pt] (0,3.7) circle (1.2);%
 \node[font=\scriptsize,inner sep=1pt] at (0,6.3) {$A$};%
 \node[font=\scriptsize,inner sep=1pt,anchor=east] at (-2.7,3.7) {$a$};%
 \node[font=\scriptsize,inner sep=1pt,anchor=west] at (2.7,3.7) {$b$};%
 \node[font=\scriptsize,inner sep=1pt,anchor=east] at (\ifnum#1=0 -2.7\else -4.3\fi,-3.7) {$a'$};%
 \node[font=\scriptsize,inner sep=1pt,anchor=west] at (\ifnum#1=0 2.7\else 4.3\fi,-3.7) {$b'$};%
 \node[font=\scriptsize,inner sep=1pt,anchor=west] at (0.4,0) {\ifnum#1=0 $s$\else $t$\fi};%
\end{tikzpicture}}}}
This appendix shows how tensor derivatives produce the open-slot contractions used in the main text. Differentiating a contraction with respect to a tensor that occurs only once removes that tensor and leaves its indices open~\cite{Liao2019}. Applied to a transfer matrix, this gives an ``open slot'', a transfer matrix with one or both tensors removed, into which other tensors can be inserted: $T(\slot,A)$ and $T(\slot,\slot)$, which Sec.~\ref{sec:energy-gradient} and Eqs.~\eqref{eq:one} and~\eqref{eq:two} use.

Consider the transfer matrix $T$ of Eq.~\eqref{eq:transfer-function}:
\begin{equation}
 T_{(aa'),(bb')}
 =\sum_s A^s_{ab}\,\overline{A^s_{a'b'}}=\osCol{0}.
 \label{eq:open-slot-T}
\end{equation}
Treating $\overline{A}$ as independent of $A$ (Wirtinger calculus), differentiating with respect to one entry of $\overline{A}$ gives
\begin{equation}
 \frac{\partial T_{(aa'),(bb')}}
 {\partial\overline{A^t_{\gamma'\delta'}}}
 =A^t_{ab}\,\delta_{a'\gamma'}\delta_{b'\delta'}=\osCol{1}.
 \label{eq:open-slot-dT}
\end{equation}
The bra tensor has disappeared, leaving its three indices $t$, $\gamma'$, and $\delta'$ open. The dashed circle marks the removed tensor, written $\slot$ in formulas. Thus the derivative is $T(\slot,A)$, and contracting its open indices with $\overline{A}$ recovers $T$. Differentiating with respect to $A$ as well removes the ket tensor at the same site, so that both tensors are removed. This gives $T(\slot,\slot)$, the open slot of the effective Hamiltonian in Sec.~\ref{sec:effective-operators}.

In a uniform iMPS the differentiated tensor occurs on every site, so the product rule sums over the positions of the slot. For example, for three transfer matrices between vectors $\bra{l}$ and $\ket{r}$ that do not depend on $A$,
\begin{equation}
 \begin{split}
  \frac{\partial}{\partial\overline{A}}\bra{l}T^3\ket{r}={}&\bra{l}T(\slot,A)\,T^2\ket{r}+\bra{l}T\,T(\slot,A)\,T\ket{r}\\
  &+\bra{l}T^2\,T(\slot,A)\ket{r}.
 \end{split}
 \label{eq:open-slot-example}
\end{equation}
This is $\bra{l}f^{[1]}[T]\bigl(T(\slot,A)\bigr)\ket{r}$ with $f(z)=z^3$, the structure of the last term of Eq.~\eqref{eq:gradient}: the derivative with respect to the shared tensor $A$ acts on every transfer matrix of the string. For a general function $f$ of $T$, the sum over the position of one inserted transfer matrix, such as $T(\slot,A)$ or $T(\slot,\slot)$ in Eq.~\eqref{eq:one}, is $f^{[1]}$ of Sec.~\ref{sec:open-sums}; the sum over the positions of two inserted transfer matrices, such as $T(\slot,A)$ and $T(A,B)$ in Eq.~\eqref{eq:two}, is $f^{[2]}$.

\section{Effective Hamiltonians in the mixed canonical form}
\label{sec:mixed}
\definecolor{hAL}{HTML}{3E9C8F}\definecolor{hALe}{HTML}{2A6E65}
\definecolor{hAR}{HTML}{B0638E}\definecolor{hARe}{HTML}{7E4465}
\newcommand{\mten}[4]{\path[fill=#1,draw=#2,line width=0.5pt] (#3,#4) circle (1.2);}
\newcommand{\mcol}[3]{\draw[hline] (#3,-2.5)--(#3,2.5);\mten{#1}{#2}{#3}{3.7}\mten{#1}{#2}{#3}{-3.7}}
\newcommand{\mcolL}[1]{\mcol{hAL}{hALe}{#1}}
\newcommand{\mcolR}[1]{\mcol{hAR}{hARe}{#1}}
\newcommand{\mixsum}{\vcenter{\hbox{\begin{tikzpicture}[x=\hu,y=\hu]
 \useasboundingbox (-5.2,-7.2) rectangle (27.6,6.6);
 \draw[hline] (0,3.7)--(22.4,3.7); \draw[hline] (0,-3.7)--(22.4,-3.7);
 \hends{3.7,-3.7}{22.4}
 \mcolL{0}\node[hop] at (0,0) {$X$};
 \mcolL{5.6}
 \draw[hline] (11.2,-2.5)--(11.2,2.5);\hhole{11.2}{3.7}\hhole{11.2}{-3.7}
 \mcolR{16.8}
 \mcolR{22.4}\node[hop] at (22.4,0) {$Y$};
 \hparl{2.9}{0}{5.6}{0.9}\hparr{8.3}{0}{5.6}{0.9}\hparl{14.1}{0}{5.6}{0.9}\hparr{19.5}{0}{5.6}{0.9}
 \node[font=\scriptsize] at (9.1,5.9) {$a$}; \node[font=\scriptsize] at (20.3,5.9) {$b$};
 \node[font=\scriptsize,text=hALe] at (2.8,-6.6) {$A_L$};
 \node[font=\scriptsize] at (11.2,-6.6) {$i$};
 \node[font=\scriptsize,text=hARe] at (19.6,-6.6) {$A_R$};
\end{tikzpicture}}}}
Section~\ref{sec:effective-operators} derived the effective Hamiltonian for the uniform tensor $A$, Eq.~\eqref{eq:one}. iDMRG and VUMPS work in the mixed canonical form, and this appendix gives the corresponding effective Hamiltonians for three centers: the center tensor $A_C$ on one site, the bond matrix $C$ between two sites, and a two-site center $A_2$. In the mixed canonical form, the tensors to the left and right of the center are the left- and right-orthonormal $A_L$ and $A_R$, which satisfy $\sum_sA_L^{s\dagger}A_L^s=\id$ and $\sum_sA_R^sA_R^{s\dagger}=\id$ and are related by the bond matrix $C$ through $A_C^s=A_L^sC=CA_R^s$~\cite{Zauner2018}. The left fixed point of $T_L=T(A_L,A_L)$ and the right fixed point of $T_R=T(A_R,A_R)$ are identities, so the norm matrix at the center is the identity and the local eigenvalue problem is a standard one. In Eq.~\eqref{eq:one}, the string to the left of the center now uses $T_L$, the string to its right uses $T_R$, and $\widetilde T_L$, $\widetilde T_R$ are their connected parts. With $a$ transfer matrices between $X$ and the center and $b$ between the center and $Y$, the ends are a distance $d=a+b+2$ apart, and the sum over $a$ and $b$ is $F_J^{[1]}$ of Eqs.~\eqref{eq:fn-notation} and~\eqref{eq:FJ1}, with $\widetilde T_L$ for the $a$ transfer matrices and $\widetilde T_R$ for the $b$ transfer matrices:
\begin{equation}
 \begin{aligned}
  H_{A_C}&=\sum_{a,b\ge0}J_{a+b+2}\,\mixsum+\cdots\\
  &=\bra{L_X}F_J^{[1]}[\widetilde T_L,\widetilde T_R]
        \bigl(T(\slot,\slot)\bigr)\ket{R_Y}+\cdots ,
 \end{aligned}
 \label{eq:heff-mixed}
\end{equation}
where $\bra{L_X}$ is built from $A_L$, $\ket{R_Y}$ from $A_R$, the projectors next to $X$ and $Y$ are left implicit in the diagram, and the dots collect the remaining terms: those with both ends on the same side of the center, those with an end on the center site, and the disconnected parts, as in Eq.~\eqref{eq:one}.

VUMPS also requires the effective Hamiltonian for the bond matrix $C$, which sits on the bond between two sites, and two-site iDMRG requires that for a two-site center $A_2$. Let $a$ and $b$ again be the numbers of transfer matrices between $X$ and the center and between the center and $Y$. The distance $d$ is one more than the number of sites between $X$ and $Y$, and this number is $a+b$ plus the number of sites that the center occupies: one for $A_C$, none for $C$, and two for $A_2$. Hence $d=a+b+2$ for $A_C$, $d=a+b+1$ for $C$, and $d=a+b+3$ for $A_2$. Since $f^{[1]}$ turns $z^n$ into the sum over the strings with $a+b=n-1$ transfer matrices around the slot, the coefficient of $z^n$ in $f(z)$ must be $J_d$ with $d=n+1$ for $A_C$, $d=n$ for $C$, and $d=n+2$ for $A_2$ ($n\ge1$). These are the coefficients of $F_J(z)$, $zF_J(z)$, and $[F_J(z)-J_1]/z$, the functions in Table~\ref{tab:slots}; their constant terms, $J_1$, $0$, and $J_2$, do not enter $f^{[1]}$. All other parts of iDMRG and VUMPS are unchanged~\cite{McCulloch2008,Zauner2018}.
\begin{table}[t]
 \caption{Open slots and generating functions for the three centers of the mixed canonical form. The columns give the open slot, the distance $d$ between the ends, and the generating function $f(z)$. The terms considered have their ends on opposite sides of the center, with $a$ transfer matrices between the left end and the center and $b$ between the center and the right end. Here $T_C(C',C)=C\otimes\overline{C'}$, and $T_2$ is the transfer matrix of a two-site tensor.}
 \label{tab:slots}
 \begin{ruledtabular}
 \begin{tabular}{llll}
  Center & Open slot & $d$ & $f(z)$\\
  \hline
  $A_C$ & $T(\slot,\slot)$ & $a+b+2$ & $F_J(z)$\\
  $C$ & $T_C(\slot,\slot)$ & $a+b+1$ & $zF_J(z)$\\
  $A_2$ & $T_2(\slot,\slot)$ & $a+b+3$ & $[F_J(z)-J_1]/z$
 \end{tabular}
 \end{ruledtabular}
\end{table}

\section{Relation to the MPO methods}
\label{sec:mpo}

In this appendix, we show for the energy that the MPO contraction is a special case of TMC and that, even when an exact MPO exists, TMC needs no more applications of the transfer matrix than the MPO contraction.

TMC evaluates the generating function $F_J(T)=\sum_{d\ge1}J_dT^{d-1}$ for arbitrary coefficients $J_d$. For a two-site interaction $J_dX_iY_{i+d}$, suppose there is an exact MPO of bond dimension $K$ that propagates intermediate states via an auxiliary matrix $M$ with entry and exit vectors $u^T$ and $v$~\cite{CrosswhiteBacon2008,Pirvu2010}, which satisfy
\begin{align}
 J_d&=u^TM^{d-1}v,\label{eq:rational-J}\\
 F_J(T)&=(u^T\otimes\id)(\id-M\otimes T)^{-1}(v\otimes\id),
 \label{eq:rational1}
\end{align}
where the geometric series converges when the spectral radius of $M$ is below one, given that $T$ has spectral radius one. The MPO contraction is therefore the TMC sum restricted to coefficients of the form~\eqref{eq:rational-J}, making the MPO construction a special case of TMC. The converse, however, does not hold: Eq.~\eqref{eq:rational-J} restricts $F_J(z)=u^T(\id-zM)^{-1}v$ to be a rational function, whereas a power-law interaction $J_d=V/d^\alpha$ yields the polylogarithm $F_J(z)=V\Li_\alpha(z)/z$, which is transcendental (Sec.~\ref{sec:calculus}) and thus admits no exact finite-bond-dimension MPO. An SOE fit with $K$ exponentials, $J_d\approx\sum_{k=1}^Kc_k\lambda_k^{d-1}$, is the rational approximation $F_J(z)\approx\sum_kc_k/(1-\lambda_kz)$. TMC evaluates such nonrational functions directly (Appendix~\ref{sec:matrix-free}), including the neutralized Coulomb case (Appendix~\ref{sec:ward}).

Importantly, this generality comes at no extra cost when an exact MPO exists: both approaches are iterative and, step by step, draw on the same Krylov vectors of $T$ [Eq.~\eqref{eq:mpo-krylov}], so TMC needs no more steps, and in each step it applies $T$ no more often than the MPO contraction. We demonstrate this below for the energy evaluation.

After splitting off the fixed-point contribution $\langle X\rangle\langle Y\rangle F_J(1)$, we compare the Krylov subspaces generated in $n_{\rm K}$ steps from $b=\ket{R_Y}$. Whereas TMC iterates the connected transfer matrix $\Tconn$ directly on $b$, the MPO contraction iterates the joint operator $M\otimes\Tconn$ on $v\otimes b$:
\begin{equation}
 \begin{aligned}
  &\mathcal K_{n_{\rm K}}(\Tconn,b)
  =\operatorname{span}\bigl\{b,\,\Tconn b,\,\dots,\,\Tconn^{n_{\rm K}-1}b\bigr\},\\
  &\mathcal K_{n_{\rm K}}(M\otimes\Tconn,v\otimes b)
  =\operatorname{span}\bigl\{v\otimes b,\,(Mv)\otimes(\Tconn b),\,\dots,\\
  &\qquad(M^{n_{\rm K}-1}v)\otimes(\Tconn^{n_{\rm K}-1}b)\bigr\}\subset\mathbb C^K\otimes\mathcal K_{n_{\rm K}}(\Tconn,b).
 \end{aligned}
 \label{eq:mpo-krylov}
\end{equation}
Both subspaces are built from the same Krylov vectors $\{b,\Tconn b,\dots,\Tconn^{n_{\rm K}-1}b\}$; the auxiliary matrix $M$ only supplies the $K$-component weight vectors $M^jv$, which contract with $u^T$ to yield $J_{j+1}=u^TM^jv$. Generating these Krylov vectors takes $n_{\rm K}-1$ applications of $T$, each of cost $O(n_{\rm ph}\chi^3)$, with $n_{\rm ph}$ the local physical dimension. Standard MPO contractions that apply $T$ to each of the $K$ auxiliary components at every step require $K(n_{\rm K}-1)$ applications of $T$, whereas TMC applies $T$ only once per step. The additional overhead of TMC---the projection $\Qproj$ and the Krylov orthogonalization at cost $O(n_{\rm K}^2\chi^2)$---is a fraction of order $n_{\rm K}/(n_{\rm ph}\chi)$ of the cost of the applications of $T$, and is therefore small for $n_{\rm K}\ll n_{\rm ph}\chi$.

Furthermore, because the MPO Krylov space lies within $\mathbb C^K\otimes\mathcal K_{n_{\rm K}}(\Tconn,b)$, TMC sacrifices no bond-space resolution. Given an orthonormal basis $U$ of $\mathcal K_{n_{\rm K}}(\Tconn,b)$, TMC operates across the entire subspace $\mathcal K_{n_{\rm K}}(\Tconn,b)$ by replacing $T$ in Eq.~\eqref{eq:rational1} with the $n_{\rm K}\times n_{\rm K}$ projected matrix $G_{n_{\rm K}}=U^\dagger\Tconn U$,
\begin{equation*}
 F_J(G_{n_{\rm K}})=(u^T\otimes\id)(\id-M\otimes G_{n_{\rm K}})^{-1}(v\otimes\id),
\end{equation*}
which is evaluated at a cost independent of $\chi$. %
\section{Matrix-free evaluation}
\label{sec:matrix-free}

This appendix describes how TMC evaluates matrix functions of $\Tconn$ using only applications of $T$ to vectors. Reference~\cite{Yang2026} writes the energy and its gradient as functions of the transfer matrix applied to vectors, evaluates them with a Krylov method that needs only applications of $T$, and uses them for gradient-based variational optimization~\cite{Hauru2021}; in the most demanding regimes, however, double-precision rounding errors limit the accuracy of this Krylov method. Surpassing this accuracy limit and accelerating the calculation require the techniques described below, which we also extend to a two-site unit cell, open-slot terms, and the energy variance.

The basic operation is applying a matrix function $f(\Tconn)$ to a source vector $b$~\cite{Saad1992,Higham2008,HighamAlMohy2010}---specifically, $\ket{R_Y}$ for $F_J(\Tconn)\ket{R_Y}$ and, by transposition, $\bra{L_X}$ for $\bra{L_X}F_J(\Tconn)$ in Eq.~\eqref{eq:energy}. Let $U_{n_{\rm K}}$ be an orthonormal basis of the Krylov subspace
\begin{equation*}
 \operatorname{span}\{b,\Tconn b,\ldots,\Tconn^{n_{\rm K}-1}b\},
\end{equation*}
with first column $b/\|b\|$, and let $G_{n_{\rm K}}=U_{n_{\rm K}}^\dagger\Tconn U_{n_{\rm K}}$ be the projected $n_{\rm K}\times n_{\rm K}$ matrix. Then
\begin{equation}
 f(\Tconn)b\simeq\|b\|\,U_{n_{\rm K}}f(G_{n_{\rm K}})e_1,
 \label{eq:krylov}
\end{equation}
where $e_1$ is the first unit vector, so that $f$ is evaluated only on the small matrix $G_{n_{\rm K}}$. Since $\Tconn=\Qproj T\Qproj$ acts on the range of $\Qproj$, each step consists of applying $T$ followed by $\Qproj$. Projecting at every step prevents rounding errors from accumulating along the dominant eigenvector $\ket{R}$, which $T$ alone does not damp because $T\ket{R}=\ket{R}$.

All Krylov vectors remain in double precision. The small projected matrix needs more care. Because $\Tconn$ is not normal, $G_{n_{\rm K}}$ can have closely spaced Ritz values and an ill-conditioned matrix of eigenvectors, and its eigendecomposition then amplifies rounding errors. In these most demanding regimes, we evaluate $f(G_{n_{\rm K}})$ through resolvents, which avoid the eigendecomposition, and in extended precision to suppress rounding errors.

For a power-law interaction, the resolvent route relies on the Laplace integral representation, valid for $\alpha>0$ and $z$ outside the real interval $[1,\infty)$,
\begin{equation}
 F_J(z)=\frac{V}{\Gamma(\alpha)}\int_0^\infty
       \frac{t^{\alpha-1}\,\dd t}{\e^t-z}.
 \label{eq:mf-integral}
\end{equation}
Equation~\eqref{eq:mf-integral} follows from $d^{-\alpha}=\Gamma(\alpha)^{-1}\int_0^\infty t^{\alpha-1}\e^{-dt}\,\dd t$ and the geometric series $\sum_{d\geq1}z^{d-1}\e^{-dt}=1/(\e^t-z)$, and for $J_d=V/d^\alpha$ it equals $V\Li_\alpha(z)/z$. The two-site unit cell inserts a factor $\e^{-\sigma t}$ in the integrand, which shifts the distance $d$ by $\sigma$ (with $\sigma\in\{\pm\tfrac12\}$ for inter-sublattice couplings). Replacing $z$ by $G_{n_{\rm K}}$ expresses $F_J(G_{n_{\rm K}})$ as an integral over the small resolvents $(\e^t-G_{n_{\rm K}})^{-1}$. Each resolvent is a linear solve, so no eigenvector of $G_{n_{\rm K}}$ is needed. We evaluate the integral using the trapezoidal rule in $\ln t$. Unlike a fixed SOE fit to $J_d$, this quadrature acts exclusively on $G_{n_{\rm K}}$; its range can therefore extend far beyond the correlation length, pushing the discretization error below working precision rather than introducing a model error.

In the effective Hamiltonian of Eq.~\eqref{eq:one}, the open slot splits each interaction string into a left part $\bra{L_X}T^a$ and a right part $T^b\ket{R_Y}$, and the weight $J_{a+b+2}$ couples $a$ and $b$, so the term is a linear map on the local tensor rather than $F_J$ applied to a single vector. We build Krylov bases from $\bra{L_X}$ and from $\ket{R_Y}$, in which this coupling is a small matrix given by the Fr\'echet derivative of $F_J$, the block-matrix function of Eq.~\eqref{eq:fn-notation}~\cite{KandolfRelton2017}. Its singular value decomposition, truncated below a relative threshold of $10^{-13}$, writes the term as a short sum $\sum_k\bra{\ell_k}T(\slot,\slot)\ket{r_k}$ of left and right environments~\cite{KandolfEtAl2021}.

Finally, the energy variance $v$ of Sec.~\ref{sec:variance-method} couples two long-range operator strings. We obtain its connected part from a linear fixed-point equation and evaluate the resulting matrix functions of the Coulomb interaction analytically on the Krylov subspace. Subtraction between large competing contributions limits the absolute accuracy of $v$ to about $10^{-13}$, which sets the numerical floor reported in Sec.~\ref{sec:coulomb}. %
\section{Ward identity and the \texorpdfstring{$1/d$}{1/r} divergence}
\label{sec:ward}
A Ward identity is the local form of a conservation law: it relates an insertion of the conserved charge to the action of the symmetry~\cite{Ward1950}. For the $\Uone$ symmetry of particle number, the iMPS tensor $A^s$ of one unit cell satisfies, for every rotation angle $\theta$,
\begin{equation}
 \e^{\ii\theta\delta n_s}A^s=\e^{-\ii\theta \hat Q}A^s\e^{\ii\theta \hat Q},
 \label{eq:ward-rotation}
\end{equation}
where $\delta n_s=\sum_{i\in\text{cell}}(n_i-\rho)$ is the deviation of the particle number of the cell in the physical state $s$ from its neutralizing background $\rho$ per site, and $\hat Q=\operatorname{diag}(q_a)$ is the virtual charge operator, whose diagonal entry $q_a$ is the $\Uone$ charge of the virtual index $a$. Its derivative at $\theta=0$ is the Ward identity: the charge inserted on the physical leg equals the difference of the virtual charges inserted on the right and left legs~\footnote{This identity also follows from the construction of $\Uone$-symmetric tensors, whose blocks obey $A^s_{ab}\neq0$ only if $q_b-q_a=\delta n_s$. The Ward identity is the more general statement; in this appendix the two origins are equivalent.},
\begin{equation}
 \delta n_sA^s=A^s\hat Q-\hat QA^s,
 \label{eq:ward-commutator}
\end{equation}
\begin{center}%
\definecolor{wbl}{HTML}{5271C4}\definecolor{wble}{HTML}{3A55A0}
\definecolor{wor}{HTML}{F6C58B}\definecolor{wore}{HTML}{C77A1E}
\tikzset{wline/.style={line width=1pt,line cap=round},
  wop/.style={draw=wore,fill=wor,rounded corners=2pt,minimum height=9.5pt,minimum width=10pt,inner sep=0.8pt,line width=0.5pt,font=\scriptsize}}
\newcommand{\wardpanel}[2]{%
  \draw[wline] (#1-20,0)--(#1+20,0); \draw[wline] (#1,0)--(#1,19);
  \path[fill=wbl,draw=wble,line width=0.5pt] (#1,0) circle (6.5); \node[white,font=\scriptsize] at (#1,0) {$A$};
  \node[font=\scriptsize,inner sep=0pt] at (#1,24) {$s$};
  \node[font=\scriptsize,inner sep=0pt] at (#1-21.5,-7) {$a$}; \node[font=\scriptsize,inner sep=0pt] at (#1+21.5,-7) {$b$};
  \ifcase #2 \node[wop] at (#1,12.5) {$\delta n$}\or \node[wop] at (#1+13,0) {$\hat Q$}\or \node[wop] at (#1-13,0) {$\hat Q$}\fi;}
\newcommand{\wardidentity}{\begin{tikzpicture}[x=1pt,y=1pt]
  \wardpanel{0}{0}\node[font=\small] at (29,0) {$=$};
  \wardpanel{58}{1}\node[font=\small] at (87,0) {$-$};
  \wardpanel{116}{2}
\end{tikzpicture}}
\wardidentity\end{center}
where the physical index $s$ points up and the virtual indices $a,b$ point left and right.

We use the Ward identity to remove the singularity of $F_J(z)=-V\ln(1-z)/z$ at the dominant eigenvalue $z=1$ of $T$ analytically. The projectors in Eq.~\eqref{eq:energy} already make the energy well defined. In floating-point arithmetic, however, the projected end vectors $\bra{L_X}$ and $\ket{R_Y}$ retain a small residual along the nondecaying modes $\bra{L}$ and $\ket{R}$. This residual enters $\bra{L_X}F_J(\Tconn)\ket{R_Y}$ and is amplified by $F_J(z)\sim-V\ln(1-z)$ as $z\to1^-$, and in the gradient by $F_J'(z)\sim V/(1-z)$.

Let $T\ket{X}=\sum_sA^sXA^{s\dagger}$ be the transfer matrix, $T^{\delta n}\ket{X}=\sum_s\delta n_s\,A^sXA^{s\dagger}$ the transfer matrix with the charge deviation inserted, and $\hat Q\ket{X}=\ket{\hat QX}$ the multiplication of the matrix $X$ by $\hat Q$ from the left. For an arbitrary vector $\ket{X}$, Eq.~\eqref{eq:ward-commutator} gives
\begin{equation}
 T^{\delta n}\ket{X}=\sum_s(A^s\hat Q-\hat QA^s)XA^{s\dagger}=\bigl(T\hat Q-\hat QT\bigr)\ket{X},
 \label{eq:ward-endpoints}
\end{equation}
so a charge deviation at the end of the interaction acts as a commutator with $\hat Q$. Both ends of the interaction are charge deviations, $X=Y=\delta n$. For the environments, with $\bra{L}T=\bra{L}$ and $T\ket{R}=\ket{R}$, Eq.~\eqref{eq:ward-endpoints} gives $\bra{L}T^{\delta n}=\bra{\hat QL}(\id-T)$ and $T^{\delta n}\ket{R}=-(\id-T)\ket{\hat QR}$. Since $\bra{L}(\id-T)=0$ and $(\id-T)\ket{R}=0$, the projector $\Qproj$ of Eq.~\eqref{eq:connected-channel} leaves these vectors unchanged, so $\bra{L_X}=\bra{\hat QL}(\id-T)$ and $\ket{R_Y}=-(\id-T)\ket{\hat QR}$.

Since $T=\ket{R}\bra{L}+\Tconn$, these vectors are $\bra{L_X}=\bra{\hat QL}P(\id-\Tconn)$ and $\ket{R_Y}=-(\id-\Tconn)P\ket{\hat QR}$. Inserting them into the connected contraction with $n$ open slots $\bm\Delta=(\Delta_1,\dots,\Delta_n)$ gives
\begin{equation}
 \begin{aligned}
 &\bra{L_X}\ddf{n}[\Tconn](\bm\Delta)\ket{R_Y}\\
 &=-\bra{\hat QL}P(\id-\Tconn)\,\ddf{n}[\Tconn](\bm\Delta)(\id-\Tconn)P\ket{\hat QR}.
 \end{aligned}
 \label{eq:ward-regularized}
\end{equation}
The two factors $\id-\Tconn$ thus add a factor $(1-z)^2$ to $\ddf{n}$. For $J_d=V/d$ the function $(1-z)^2F_J(z)$ is finite at $z=1$, since $(1-z)^2\ln(1-z)\to0$, and its coefficient of $z^{d-1}$ is, for $d\geq3$, the second difference $J_d-2J_{d-1}+J_{d-2}=2V/[(d-2)(d-1)d]$, which decays as $d^{-3}$. This cancels the divergence of $\ddf{n}$ at $z=1$, so that a component along $\ket{R}$ built up by rounding errors in $\Tconn$ is no longer amplified.

In the first-quantized iMPS of Sec.~\ref{sec:firstq}, the background is the term $-\rho$ of Eq.~\eqref{eq:firstq-eV}, and for $\alpha=1$ it cancels the divergence of the disconnected term of Eq.~\eqref{eq:firstq-split}. Near $\alpha=1$ the Hurwitz zeta function behaves as $\zeta(\alpha,\eta)=1/(\alpha-1)-\psi(\eta)+O(\alpha-1)$, where $\psi$ is the digamma function, so its pole does not depend on $\eta$. With $\sum_gw_g=1$, the poles of $F_{\alpha,\Delta}(1)=\mu^{-\alpha}\zeta(\alpha,s_0+\Delta/\mu)$ in the disconnected term cancel that of the background, $-\rho\,\zeta(\alpha,1)$, and the limit $\alpha\to1$ gives the finite value $-\rho\bigl[\sum_{g,g'}w_gw_{g'}\,\psi\bigl(s_0+(g'-g)/\mu\bigr)+\ln\mu+\gamma_{\rm E}\bigr]$ in $e_V/V$, with Euler's constant $\gamma_{\rm E}=-\psi(1)$.
   
\section{Finite-size energy benchmark at \texorpdfstring{$\alpha=1$}{alpha=1}}
\label{sec:ed-alpha-one}
This appendix checks the iMPS energies of Sec.~\ref{sec:coulomb} by ED of finite chains.

\begin{table}[!b]
\centering\footnotesize
\renewcommand{\arraystretch}{1.18}
\begin{tabular*}{\columnwidth}{@{\extracolsep{\fill}}cccccc}
\toprule
 & \multicolumn{2}{c}{OBC} & \multicolumn{2}{c}{PBC} & \\
\cmidrule(lr){2-3}\cmidrule(lr){4-5}
$V$ & $L^{-1}$ & $+L^{-2}$ & $L^{-2}$ & $+L^{-4}$ & This work\\
\midrule
2 & $-0.86327$ & $-0.86375$ & $-0.86402$ & $-0.86391$ & $-0.86382$\\
4 & $-1.11121$ & $-1.11190$ & $-1.11235$ & $-1.11220$ & $-1.11210$\\
\bottomrule
\end{tabular*}
\caption{Finite-size benchmark of the Coulomb-chain energies. The table lists the energies per site in the thermodynamic limit obtained from the four ED extrapolations and from the zero-variance iMPS extrapolations (Table~\ref{tab:benchmark}). The columns headed $+L^{-2}$ and $+L^{-4}$ give the energies extrapolated from fits that include the subleading terms $b/L^2$ for OBC and $b/L^4$ for PBC, respectively.}
\label{tab:ed-alpha-one}
\end{table}

For this benchmark, we diagonalize the Hamiltonian of $N=L/2$ particles on $L$ sites,
\begin{equation}
 H_L=-\sum_{i}(c_i^\dagger c_{i+1}+\mathrm{h.c.})+V\sum_{i<j}\frac{(n_i-1/2)(n_j-1/2)}{d_L(i,j)},
\end{equation}
with $d_L(i,j)=|i-j|$ for open boundary conditions (OBC, $L=6,8,\ldots,30$) and the chord distance $d_L(i,j)=(L/\pi)\sin(\pi|i-j|/L)$ for periodic boundary conditions (PBC, $L=6,10,\ldots,38$, for which $N$ is odd). We use reflection and particle-hole symmetry, together with translation symmetry for PBC, to reduce the dimension of the Hamiltonian matrix to be diagonalized. We then compute the ground state with ARPACK~\cite{Lehoucq1998ARPACK}.

We fit $e(L)=E_0(L)/L$ to $e_\infty+a/L$ (with an optional $+b/L^2$ term) for OBC and to $e_\infty+a/L^2$ (with an optional $+b/L^4$ term) for PBC. For each boundary condition, we use unweighted least squares over the six largest sizes ($L\geq20$ for OBC and $L\geq18$ for PBC), both with and without the subleading term.

The OBC and PBC extrapolations bracket the iMPS energies and, with the subleading terms, agree with them to within $2.1\times10^{-4}$ (Table~\ref{tab:ed-alpha-one}), the accuracy of the finite-size extrapolation. 
\makeatletter
\immediate\write\@auxout{\string\citation{prxqBibliographyControl}}
\makeatother
\bibliographystyle{apsrev4-2}
\bibliography{references}
\end{document}